\documentclass[letterpaper,twocolumn,10pt]{article}
\usepackage{newStyle}
\usepackage{tikz}
\usepackage{amsmath}

\usepackage{filecontents}

\usepackage{enumitem}
\usepackage{balance}
\usepackage{booktabs}  
\usepackage{algorithmic}
\usepackage{graphicx}
\usepackage{textcomp}
\usepackage{xcolor}
\usepackage{xspace}
\usepackage{subfigure}
\usepackage{array}
\usepackage{makecell}
\usepackage{multirow}
\usepackage{hyperref}

\hypersetup{
  colorlinks   = true,    % Colours links instead of ugly boxes
  urlcolor     = blue,    % Colour for external hyperlinks
  linkcolor    = blue,    % Colour of internal links
  citecolor    = red      % Colour of citations
}

\usepackage[utf8]{inputenc} % allow utf-8 input
\usepackage[T1]{fontenc}    % use 8-bit T1 fonts
\usepackage{booktabs}       % professional-quality tables
\usepackage{amsmath}

\usepackage{amsfonts}       % blackboard math symbols
\usepackage{nicefrac}       % compact symbols for 1/2, etc.
\usepackage{microtype}      % microtypography
\usepackage{subfigure}
\usepackage{graphicx}
\graphicspath{{figures/}}
\usepackage{array}
\usepackage{makecell}
\usepackage[ruled,vlined]{algorithm2e}  % linesnumbered
\usepackage{multirow}

\usepackage{diagbox}

\usepackage{xcolor}

\usepackage{cleveref}

\usepackage{xspace}
\usepackage{cleveref}
\newcommand{\sysname}{\texttt{Aegis}\xspace}
\newcommand{\desdn}{\texttt{DESDN}\xspace}
\newcommand{\rob}{\texttt{ROB}\xspace}

\newcommand{\wjl}[1]{\textcolor{red}{#1}}

\usepackage{flushend}

\newcommand{\PreserveBackslash}[1]{\let\temp=\\#1\let\\=\temp}
\newcolumntype{C}[1]{>{\PreserveBackslash\centering}p{#1}}
\newcolumntype{R}[1]{>{\PreserveBackslash\raggedleft}p{#1}}
\newcolumntype{L}[1]{>{\PreserveBackslash\raggedright}p{#1}}

\usepackage{tikz}

\usepackage[misc]{ifsym}

\makeatletter
\newcommand{\removelatexerror}{\let\@latex@error\@gobble}
\makeatother
\definecolor{clthu}{RGB}{85, 43, 111}
\definecolor{cluci}{RGB}{219, 109, 0}
\definecolor{clqianxin}{RGB}{60, 174, 225}
\definecolor{clzgc}{RGB}{226, 32, 17}
\newcommand{\mkthu}[0]{{{$^*$}}}
\newcommand{\mkbupt}[0]{{{$^\dag$}}}
\newcommand{\mkntu}[0]{{{$^\ddag$}}}
\newcommand{\mkzgc}[0]{{{$^\S$}}}
\newcommand{\mkant}[0]{$^	\spadesuit$}
\newcommand{\mkhd}[0]{$^	\clubsuit$}
\newcommand{\mkletter}[0]{{{\normalsize \textsuperscript{\Letter}}}}
\renewcommand\footnotemark{}

\usepackage{xspace}
\usepackage{amsmath}
\usepackage{amssymb}
\usepackage{bm}
\usepackage[ruled,vlined]{algorithm2e}
\newenvironment{packeditemize}{
\begin{list}{$\bullet$}{
\setlength{\labelwidth}{8pt}
\setlength{\itemsep}{0pt}
\setlength{\leftmargin}{\labelwidth}
\addtolength{\leftmargin}{\labelsep}
\setlength{\parindent}{0pt}
\setlength{\listparindent}{\parindent}
\setlength{\parsep}{0pt}
\setlength{\topsep}{3pt}}}{\end{list}}

\begin{document}

\title{\sysname: Mitigating Targeted Bit-flip Attacks against Deep Neural Networks}

\author{
    {\rm Jialai Wang}\mkthu \rm ,
    {\rm Ziyuan Zhang}\mkbupt \rm ,
    {\rm Meiqi Wang}\mkthu \rm ,
    {\rm Han Qiu}\mkthu \mkzgc \mkletter\thanks{\mkletter~Corresponding authors.} \rm ,
    {\rm Tianwei Zhang}\mkntu \rm ,\\
    {\rm Qi Li}\mkthu \mkzgc \rm ,
    {\rm Zongpeng Li}\mkthu \mkhd
\mkletter
    \rm ,
    {\rm Tao Wei}\mkant \rm ,
    {\rm and Chao Zhang}\mkthu \mkzgc\rm
    \\
    \mkthu {Tsinghua University},
    \mkbupt {Beijing University of Posts and Telecommunications},\\
    \mkntu {Nanyang Technological University},
    \mkhd {Hangzhou Dianzi University},
    \mkant {Ant Group},\\
    \mkzgc Zhongguancun Laboratory
    % \mkpc \usenixhref{https://www.pcl.ac.cn/}{Peng Cheng Laboratory}
    \medskip
    \\
    \{{wang-jl22, wang-mq22}\}@mails.tsinghua.edu.cn,
    zhangziy0421@bupt.edu.cn, \\
    \{{qiuhan, qli01, zongpeng, chaoz}\}@tsinghua.edu.cn,
    tianwei.zhang@ntu.edu.sg,\\
    lenx.wei@antgroup.com
}

\maketitle
% \author{{\rm Jialai Wang}
% \\
% {Tsinghua University, BNRist}
% \\
% {\rm wang-jl22@mails.tsinghua.edu.cn}
% \and
% {\rm Jialai Wang}
% \\
% {Tsinghua University, BNRist}
% \\
% {\rm wang-jl22@mails.tsinghua.edu.cn}
% \and
% {\rm Jialai Wang}
% \\
% {Tsinghua University, BNRist}
% \\
% {\rm wang-jl22@mails.tsinghua.edu.cn}
% % \authornote{Institute for Network Sciences and Cyberspace, Tsinghua University}
% }

% \affiliation{%
%   \institution{Tsinghua University, BNRist}
%   \city{Beijing}
%   \country{China}
% }
% \email{wangjl19@mails.tsinghua.edu.cn}

\begin{abstract}
Bit-flip attacks (BFAs) have attracted substantial attention recently, in which an adversary could tamper with a small number of model parameter bits to break the integrity of DNNs.
To mitigate such threats, a batch of defense methods are proposed, focusing on the untargeted scenarios. Unfortunately, they either require extra trustworthy applications or make models more vulnerable to targeted BFAs. 
Countermeasures against targeted BFAs, stealthier and more purposeful by nature, are far from well established. 

In this work, we propose \sysname, a novel defense method to mitigate targeted BFAs.
The core observation is that existing targeted attacks focus on flipping critical bits in certain important layers. 
Thus, we design a dynamic-exit mechanism to attach extra internal classifiers (ICs) to hidden layers. 
This mechanism enables input samples to early-exit from different layers, which effectively upsets the adversary's attack plans. 
Moreover, the dynamic-exit mechanism randomly selects ICs for predictions during each inference to significantly increase the attack cost for the adaptive attacks where all defense mechanisms are transparent to the adversary.
We further propose a robustness training strategy to adapt ICs to the attack scenarios by simulating BFAs during the IC training phase, to increase model robustness. 
Extensive evaluations over four well-known datasets and two popular DNN structures reveal that \sysname could effectively mitigate different state-of-the-art targeted attacks, reducing attack success rate by 5-10$\times$, significantly outperforming existing defense methods. 
We open source the code of \sysname\footnote{{https://github.com/wjl123wjl/Aegis.git}}.
\end{abstract}

\section{Introduction}

The recent revolutionary development of deep neural network (DNN) models has promoted various security- and safety-sensitive intelligent applications, such as autonomous driving~\cite{DBLP:conf/uss/Sato0WJLC21}, AI on satellites~\cite{ghiglione2022opportunities}, and medical diagnostics~\cite{shickel2017deep}.  
An adversary could manipulate data used by DNN models and model parameters to launch various attacks.
The security and robustness of DNN models have become the key factors affecting the deployment of these systems. 
A significant amount of research efforts have been devoted to protecting DNN models from data-oriented attacks, e.g. adversarial attacks~\cite{DBLP:journals/corr/GoodfellowSS14,carlini2017towards,madry2018towards,fnu2020hybrid,DBLP:conf/uss/LovisottoTSSM21,DBLP:conf/uss/HussainNDMK21} that manipulate inference data, or DNN backdoor attacks~\cite{adi2018turning,li2020backdoor,DBLP:conf/uss/SeveriMCO21,liu2020reflection,DBLP:conf/iccv/LiLWLHL21,DBLP:conf/cvpr/ZhaoMZ0CJ20,DBLP:conf/emnlp/LiSLZMQ21,dong2022mind} that manipulate training data. 
These efforts can secure the model against data-oriented threats. 
But little attention has been paid to mitigate the emerging parameter-oriented attacks.

Recent studies have shown that well-trained DNN models are vulnerable to parameter-oriented attacks, which tamper with model parameters~\cite{rakin2021t,DBLP:conf/cvpr/RakinHF20,DBLP:conf/iccv/RakinHF19,DBLP:conf/iccv/ChenFZK21,yao2020deephammer,DBLP:conf/uss/RakinLXF21}. 
For instance, flipping a small number of critical bits (i.e. $0 \to 1$ or $1 \to 0$) of off-the-shelf DNN model parameters can trigger catastrophic changes in the inference process~\cite{hong2019terminal,DBLP:conf/iccv/RakinHF19}, lowering the prediction accuracy or manipulating the inference to any target labels. 
These bit-flip attacks (BFAs) experiment in real-world scenarios.  DeepHammer~\cite{yao2020deephammer} performs BFAs on a PC via rowhammer. 
Also, BFAs are performed on the multi-tenant FPGA devices in cloud-based machine learning services~\cite{DBLP:conf/uss/RakinLXF21}. 

Current state-of-the-art BFAs can be classified into untargeted and targeted attacks. 
The \textit{untargeted} BFAs aim to comprise the victim model accuracy to the random guess level~\cite{DBLP:conf/iccv/RakinHF19,yao2020deephammer,DBLP:conf/uss/RakinLXF21}. 
For instance, with optimized critical bit search algorithms, the BFA in~\cite{DBLP:conf/iccv/RakinHF19} needs to flip only 13 out of 93 million bits of an 8-bit quantized ResNet-18 model on ImageNet, to degrade its top-1 accuracy from 69.8\% to 0.1\%.
In comparison, the \textit{targeted} BFAs are stealthier, by misleading target models to target labels on specific samples (i.e., sample-wise attacks) or samples with special triggers (i.e., backdoor attacks) while preserving model accuracy for other samples. For instance, the sample-wise BFA~\cite{ DBLP:conf/iclr/BaiWZL0X21} can flip only less than 8 critical bits on average of an 8-bit quantized ResNet-18 model on ImageNet to manipulate the prediction of specific input samples.

Existing defenses for data-oriented attacks, e.g., adversarial training, are proven not useful to mitigate BFAs~\cite{DBLP:conf/cvpr/HeRLCF20}.  
Instead, a small number of dedicated defenses have been proposed to mitigate BFAs, which can be classified into two categories, namely integrity verification, and model enhancement.

First, integrity verification-based approaches~\cite{DBLP:conf/iccad/JavaheripiK21,li2020deepdyve,DBLP:conf/iccad/LiuWW20} verify the integrity of model parameters at runtime to detect BFAs.
For instance, HashTAG~\cite{DBLP:conf/iccad/JavaheripiK21} verifies the integrity of model parameters on the fly by extracting and comparing the unique signatures of the original and the runtime DNN models. 
A low-collision hashing scheme could be used for generating signatures and achieving almost zero false positives. 
This type of approach could detect any attempts to tamper with the models and thus could defeat both targeted and untargeted BFAs.
However, they in general require an additional \textit{trusted} and \textit{secure} monitoring process to \textit{continuously} monitor the target DNN model.
It introduces additional performance overhead and resource cost and is not applicable to commercial off-the-shelf devices and practical scenarios.

Second, model enhancement-based approaches~\cite{DBLP:conf/dac/LiRXCHFC20,zhan2021improving} focus on improving the robustness of target models directly, making BFAs difficult or impossible to launch. 
Note that, precisely flipping a large number of bits via hardware-level attack is not practical~\cite{yao2020deephammer}. 
An important metric for evaluating the difficulty or cost of BFAs is \textit{the number of bits to flip} (e.g. DeepHammer~\cite{yao2020deephammer} sets 24 bits as the maximum number of bits can be flipped). 
Thus, this approach aims to significantly increase the number of bits to flip for achieving the same attack goals. 
One promising approach is to quantize the model parameters to constrain the weights' value range that may potentially be changed by BFAs. 
For instance, a binarization architecture, BNN,  is proposed in~\cite{DBLP:conf/cvpr/HeRLCF20} to retrain the model from scratch to generate a model with weights only equal to -1 or +1. 
An enhanced version, RA-BNN~\cite{rakin2021ra} further extends the quantization on the activation function outputs by -1 or +1. 
This approach can significantly increase the difficulty for untargeted attacks (e.g. flipping $40 \sim 300\times$ more bits in~\cite{DBLP:conf/cvpr/HeRLCF20}, which is infeasible). 
However, this solution can only mitigate untargeted attacks, while making the model even more vulnerable to targeted attacks. 
According to our experiments, TBT attacks~\cite{DBLP:conf/cvpr/RakinHF20} can achieve a similar success rate by flipping even fewer bits in a binarization-trained model (from 206 bits in the vanilla model to 50 bits in the corresponding binarization model. See details in Section~\ref{exp:no-tbt}). 

In this work, we propose \sysname\footnote{\sysname is a powerful shield carried by Athena and Zeus in Greek mythology, which can defeat various attacks from thunders, hammers, etc.}, a novel method to mitigate different targeted BFAs. 
Our key observation is that existing targeted BFAs achieve their goals by locating the most critical bits according to the model inference process. 
Particularly, they flip the critical bits in either the final layer of the target model or the most important layer determined by some optimization methods. Based on the observation, we design our solution by using a \textit{dynamic multi-exit} architecture to train extra internal classifiers (ICs) for hidden layers~\cite{kaya2019shallow}. 
These ICs can distribute the early exit of input samples to different hidden layers of the target model. 
This can mitigate the existing attacks which flip bits in one specific layer. 
Furthermore, considering the adaptive BFAs where the defense is transparent, adversaries could use the sample exit distribution to locate critical hidden layers to flip their critical bits.
\sysname can mitigate such adaptive BFAs by dynamic masking certain exits.
Lastly, we design robust training on ICs to simulate the attacks when the critical bits in each hidden layer are flipped. 
This can defeat more sophisticated adaptive BFAs that include all exits to flip critical bits in all layers of the model. 

\sysname aims to achieve three important goals, i.e. \textit{non-intrusive}, \textit{platform-independent}, and \textit{utility-preserving}. 
It can protect the model without modifying any of its parameters. This will exclude the defenses that retrain the model from scratch, which is either too costly on a large-scale dataset or infeasible when training datasets are unavailable. 
We design \sysname at the application level without requiring any additional reliable program or hardware protection, making it generally applicable. 
For utility-preserving, \sysname has negligible impacts on the prediction accuracy.

We conduct extensive experiments to evaluate \sysname against three state-of-the-art targeted BFAs with different goals, as well as their potential adaptive attack counterparts. We consider two well-known model architectures (ResNet-32 and VGG-16) and four datasets (CIFAR-10, CIFAR-100, STL-10, and Tiny-ImageNet). 
The results show that we can mitigate different BFAs by significantly increasing the number of bits to flip (e.g. flip $35\times$ more bits to achieve a similar attack success rate of~\cite{DBLP:conf/iccv/ChenFZK21}) or reducing their attack success rate to a low level (e.g. keep attack success rate lower than 4\% with a similar number of bits flipped as~\cite{DBLP:conf/iclr/BaiWZL0X21}). 

\iffalse
%can-be-added-back
In summary, we make the following contributions. 

\begin{packeditemize}
    \item We propose a novel mitigation solution against different targeted BFAs by efficiently enhancing the target model with little influence on its accuracy. 
    \item We consider potential adaptive BFAs originating from existing BFAs in a white-box scenario, and prove that our solution can still mitigate these advanced threats by significantly increasing the attack difficulty. 
    \item We conduct comprehensive evaluations with two well-known model architectures and four datasets and show our solution can outperform existing works.
\end{packeditemize}
\fi

\section{Background}

\subsection{Targeted Bit-flip Attacks}
\label{background:Targeted}

We introduce three state-of-the-art targeted BFAs, i.e., TBT attack~\cite{DBLP:conf/cvpr/RakinHF20}, ProFlip attack~\cite{DBLP:conf/iccv/ChenFZK21} and TA-LBF attack~\cite{DBLP:conf/iclr/BaiWZL0X21}.

%%\vspace{-10pt}
\noindent\textbf{TBT}~\cite{DBLP:conf/cvpr/RakinHF20} is a targeted BFA that injects backdoors into the target model through flipping bits.
The attacker's goal is that the compromised model still operates with normal inference accuracy on benign inputs but makes mistakes on samples with specific triggers.
Specifically, when the adversary embeds the trigger into any input, the model is forced to classify this input to a certain target class.
Note that this method only flips bits in the final layer of the target model.
In the final layer, the adversary first selects $w_b$ critical network neurons which have the most significant impact on the target class, then generates a specific trigger to activate these neurons.
Finally, the adversary formalizes an optimization problem to modify critical bits corresponding to these neurons.

%\vspace{-10pt}
\noindent\textbf{ProFlip} ~\cite{DBLP:conf/iccv/ChenFZK21} inserts a backdoor into the target model by flipping bits in the network weights to manipulate the prediction of all inputs attached with the trigger to a certain target class.
This method could flip bits in all the layers of the model by selecting salient neurons through forwarding derivative-based saliency map construction (also known as jacobian saliency map attack (JSMA)~\cite{DBLP:conf/eurosp/PapernotMJFCS16}). 
Then the adversary uses the gradient descent method to generate triggers, which can stimulate salient neurons to large values. 
Finally, ProFlip proposes an efficient retrieval algorithm to select the optimal parameter, and determine critical bits in the parameter to flip.

%%\vspace{-10pt}
\noindent\textbf{TA-LBF}~\cite{DBLP:conf/iclr/BaiWZL0X21} does not need a trigger but only misclassifies a specific sample to a target class by flipping the critical bits of the parameters, which makes the attack stealthier than TBT and ProFlip.
The adversary formalizes the attack as binary integer programming since the parameters are stored as binary bits (i.e., 0 and 1) in the memory.
It further equivalently reformulates this binary integer programming problem as a continuous optimization problem.
Using the alternating direction method of multipliers (ADMM) method~\cite{DBLP:journals/pami/WuG19}
solves the optimization problem to determine critical bits to flip.

\subsection{Existing Defense and Analysis}
\label{sec:background-defense}
Existing defense methods could be categorized into two types. 
Details are given as follows.

\noindent\textbf{Model enhancement.} 
\sysname also falls into this defense category. 
Li et al.~\cite{DBLP:conf/dac/LiRXCHFC20} adopt a weight reconstruction method, which could defuse the changed values on several parameters to multiple parameters, thus mitigating the effects brought by untargeted BFAs.
Zhan et al.~\cite{zhan2021improving} modify rectified linear unit (ReLU), a commonly used activation function in DNNs, to tolerate the faults incurred by bit-flipping on weights.

The above two defense methods are proved to be less effective than the binarization strategy such as BIN~\cite{DBLP:conf/cvpr/HeRLCF20} and RA-BNN~\cite{rakin2021ra}.
This strategy applies the binarization-aware training~\cite{DBLP:conf/eccv/RastegariORF16} to retrain a binarization model from scratch to mitigate untargeted BFAs. 
Its point is to constrain the range of parameters' values to force attackers to flip more bits in order to achieve the same attack success rate. 
Specifically, BIN~\cite{DBLP:conf/cvpr/HeRLCF20} converts a part of the model parameters from high precision, e.g., 32-bit floating-point, to a binary format ($\{-1,+1\}$). RA-BNN ~\cite{rakin2021ra} uses a more aggressive way to further quantize the output of activation functions to $\{-1,+1\}$ as well. 
Although these methods can effectively mitigate untargeted attacks, they still have three limitations. 
First, they require retraining a target model from scratch, which introduces significant computation costs. 
Second, aggressive precision reduction on models will affect model accuracy. 
Third, more importantly, they make the model even more vulnerable to targeted attacks such as TBT~\cite{DBLP:conf/cvpr/RakinHF20} (See Section~\ref{exp:no-tbt}).

\iffalse

Specifically, this approach converts the model parameters from high precision, e.g., 32-bit floating-point, to a binary format ($\{-1,+1\}$) encoded by 1-bit.
Such precision degradation constrains the range of parameter values of BFAs
makes the method effectively mitigate untargeted BFAs. 
However, due to the aggressive model capacity reduction (each parameter occupies only 1 bit), the model accuracy is hugely affected.
~\wjl{Rakin et al.~\cite{rakin2021ra} propose RA-BNN to further enhance the robustness of the weight binarization solution~\cite{DBLP:conf/cvpr/HeRLCF20}. They binarize both the weights and the activations to mitigate BFAs and increase the number of channels per convolutional layer for model accuracy.
However, the activations of the fully connected layer (final layer) are not binarized.
Therefore, targeted BFAs flipping bits in the final layer still threaten RA-BNN.  
Besides, both RA-BNN and BIN are not lightweight, as they require retraining the whole model from scratch.
This limits the practicality of these defense methods, as commercial models are expensive to retrain in the real world.
}
\fi

%\vspace{-10pt}
\noindent\textbf{Integrity verification.}
This approach is orthogonal to model enhancement, which protects models from another dimension. 
One approach~\cite{DBLP:conf/iccad/JavaheripiK21,DBLP:conf/iccd/GuoLCZY21,DBLP:conf/date/LiRHFC21,DBLP:conf/iccad/LiuWW20} is to apply the integrity verification to defend BFAs is that the defender extracts a ground-truth signature from the model before deployment.
Once the model is deployed, new hashes are extracted during inference to compare with the ground-truth one. 

This approach can also be realized at the hardware or system level based on techniques such as ECC~\cite{DBLP:conf/sp/CojocarRGB19}. 
However, it has three main practical obstacles. 
(1) They are restricted to specific platforms, e.g.,  \cite{DBLP:conf/iccd/GuoLCZY21} requires new CUDA kernel for integrity protection and~\cite{jiang2021trrscope} requires new processors with targeted row refresh. 
Also, some techniques such as ECC are not deployed in some embedded devices such as Nvidia Nano or Jetson AGX Xavier. 
(2) These methods are not absolutely secure against bit-flip attack~\cite{DBLP:conf/sp/CojocarRGB19}. 
(3) They only detect whether a model is changed in memory, but do not provide mitigation against specific attacks.

\noindent\textbf{Comparison with existing defense.} 
Existing model enhancement methods could effectively mitigate untargeted BFAs, but pay no attention to targeted BFAs.
Compared with untargeted attacks, targeted attacks are more threatening and stealthier, as the compromised model could still behave normally on clean samples.
Thus, we aim to fill this gap. Besides, \sysname is non-intrusive compared with existing methods as we do not modify the original models or retrain from scratch.

Integrity verification approaches experiments via hardware or system-level solutions. 
Instead, \sysname aims to give an application-level solution that is generally effective regardless of the underlying hardware circuits, operating systems, or DL libraries. 
Besides, \sysname is orthogonal to integrity verification such as ECC so \sysname can provide extra protection on ECC-enabled systems on different levels. 

We also notice there are defense methods that are specific to DNN backdoor attacks~\cite{DBLP:conf/eccv/WangZLCXW20,DBLP:conf/acsac/GaoXW0RN19,DBLP:conf/iclr/ZengCPM0J22,DBLP:conf/iclr/LiLKLLM21}. 
However, they have different threat models with mitigating BFAs. 
Specifically, they aim to detect or remove an existing backdoor in offline trojan models. 
However, BFAs usually experiment on a deployed clean model under attack at runtime.

\subsection{Multi-exit DNN models}
\label{sec:background-sdn}

\iffalse
\begin{figure}[htbp]
    \centering
 \includegraphics[width=0.4\textwidth,trim=0.0cm 5cm 4.8cm 0cm,clip
    ]{figures/sdn} 
    \vspace{-2ex}
    \caption{Deploying SDN in resource-constrained scenarios to distribute inference on the edge device and cloud server.}
    \label{fig:sdn}
    \vspace{-3ex}
% \setlength{\belowcaptionskip}{-1.5cm}
\end{figure}
\fi

The initial motivation for setting exits for inference at hidden layers is to solve the overthinking issue. 
Since the growing performance of modern DNN models brings a significantly increasing number of layers and parameters in most of the state-of-the-art DNN models, 
Huang \textit{et al.} \cite{huang2017multi} point out that forcing all samples, especially canonical samples to inference all layers of a DNN model definitely brings a waste of energy and time. 
Moreover, Kaya \textit{et al.} ~\cite{kaya2019shallow} find that forcing certain samples classified correctly with only a few shallow layers but inference through all layers will lead to the wrong prediction. 

Many solutions have been proposed to let samples early exit the model to address the above issues~\cite{wang2018skipnet,huang2017multi,he2019model,he2020attacking}. 
One promising technique is the shallow-deep network (SDN)~\cite{kaya2019shallow}. 
The key insight of SDN is that during the inference process for a sample, it is highly possible that some layer in the middle of the network already has high confidence for prediction. 
So it can early exit from the model without the need to go through all the layers to significantly reduce the inference time and energy consumption. 
It is very convenient to convert a vanilla DNN model (e.g. ResNet) into an SDN model. 
We can select some appropriate convolution layers, and attach an internal classifier (IC) to each of them to form an early exit. 
When the prediction confidence of the input sample as one label is higher than a threshold at an exit, the inference will stop and output that label. 
A proper threshold can realize early-exit with a tiny accuracy loss.

Deploying multi-exit model architectures such as SDN for security purposes is proposed in~\cite{DBLP:conf/iclr/HuCWW20,zhou2020bert} to mitigate adversarial attacks. 
Particularly, an input-adaptive multi-exit DNN structure with a dynamic inference process can mitigate the adversarial perturbation generation and further increase the difficulty of adaptive adversarial attacks. 
However, since BFAs aim at manipulating model parameters rather than input samples, simply deploying multi-exit DNN structures cannot achieve the defense requirements and more sophisticated methods are needed.

\section{Threat Model and Defense Requirements}
\label{sec:theatModel}

We consider an adversarial scenario, where the adversary is able to perform BFAs against the victim DNN models. 
He can precisely flip a number of parameter bits to affect the model prediction results. 
The exploitability and practicality of such threats have been validated and evaluated in previous works~\cite{yao2020deephammer,DBLP:conf/uss/RakinLXF21,breier2018practical,DBLP:conf/sp/RakinCYF22}. 
For instance, attackers verified an untargeted BFA on DNNs with a PC platform via row-hammer~\cite{DBLP:conf/uss/RakinLXF21}. 
Also, the adversary can co-locate his malicious program on the same machine with the victim DNN model and then use methods like row-hammer~\cite{DBLP:conf/uss/RakinLXF21} to perform BFAs. 
The feasibility of BFAs is also validated in our paper on a PC platform following previous works~\cite{yao2020deephammer,DBLP:conf/sp/RakinCYF22} (see details in Section~\ref{sec:feasibility}).

Following the previous works~\cite{yao2020deephammer,DBLP:conf/cvpr/RakinHF20,DBLP:conf/iclr/BaiWZL0X21,DBLP:conf/iccv/ChenFZK21,DBLP:conf/sp/RakinCYF22}, we assume the adversary has very strong capabilities. 
He has full knowledge of the victim model, including the DNN architecture, model parameters, etc. 
We further assume that the adversary knows every detail of any possible defense deployed in the system, such as the mechanism, algorithm, parameters, etc. 
If a defense solution employs randomization-based techniques, we assume the random numbers generated in real-time are perfect with a large entropy such that the adversary cannot obtain or guess the correct values. 
It is worth noting that these assumptions represent the strongest adversary, which significantly increases the difficulty of defense designs.

\noindent\textbf{Adversarial goals.}
Previous works have demonstrated different goals for BFAs, as summarized below:

\begin{packeditemize}
    \item \textbf{Untargeted attack}. The adversary aims to drastically degrade the overall accuracy of the victim model. A powerful untargeted attack can decrease the model accuracy to nearly random guess after the exploitation~\cite{yao2020deephammer}.
    \item \textbf{Backdoor targeted attack}. The adversary designs a specific trigger and injects the corresponding backdoor \cite{DBLP:conf/iccv/ChenFZK21} into the DNN via the BFAs. Then for any input sample containing the trigger, the compromised model will mispredict it as the target class. 
    \item \textbf{Sample-wise targeted attack}. The adversary aims to tamper with the model such that it only mispredicts a specific sample as the target class while having normal predictions for other samples. 
\end{packeditemize}

In this paper, we focus on the last two categories of targeted BFAs due to two reasons. (1) As indicated in Section \ref{sec:background-defense}, a number of past works have explored the mitigation approaches against untargeted BFAs. In contrast, how to effectively thwart targeted BFAs is rarely investigated. (2) The backdoor or sample-wise targeted BFAs are much stealthier than the untargeted attack, as the compromised model behaves normally for clean samples. This significantly increases the defense difficulty and an effective solution is urgently needed. 
Besides, although untargeted BFAs are not within our scope, our approach can mitigate untargeted BFAs (see Section~\ref{sec:discussion}).

\noindent\textbf{Defense requirements.}
The purpose of this paper is to design a defense approach to comprehensively protect DNN models from different targeted BFAs and their possible adaptive attacks. It is worth highlighting that our goal is to increase the attack cost rather than totally preventing BFAs. Theoretically, the adversary can tamper with more parameter bits even if a strong defense is applied. So we aim to significantly increase the number of flipped bits required to achieve the desired adversarial goal, thus making the attack less feasible or practical. Our defense requirements are as follows.

\begin{packeditemize}
    \item 
    \textbf{Non-intrusive}. The defense should be easy to deploy on off-the-shelf DNN models. The defender does not need to modify parameters of the original model, e.g., retraining a model with binarization~\cite{DBLP:conf/cvpr/HeRLCF20} from scratch since this can incur significant computation cost, especially for large-scale DNN models (e.g. ImageNet scale~\cite{DBLP:journals/ijcv/RussakovskyDSKS15}). 

    \item \textbf{Platform-independent}. Previous works propose hardware or system-level solutions to prevent fault injection attacks, e.g., new CUDA kernels for integrity protection~\cite{DBLP:conf/iccd/GuoLCZY21}, new processors with targeted row refresh \cite{jiang2021trrscope}. However, these solutions are restricted to some specific platforms. Instead, we hope to have an application-level solution that is generally effective regardless of the underlying hardware circuits, operating systems, and deep learning libraries. 
    \item \textbf{Utility-preserving}. The defense solution should have a negligible impact on the model inference process. It should preserve the usability of the original model without hugely decreasing its prediction accuracy. 

\end{packeditemize}

\section{Methodology}

\subsection{Design Insight}

We propose \sysname, a novel approach to mitigate different types of targeted BFAs. 
Our approach is composed of a Dynamic-Exit SDN (\desdn) mechanism followed by a robust training (\rob) strategy only on ICs. 
We illustrate our design insight via three steps as follows.

\begin{figure}[!h]
    \centering
 \includegraphics[width=.48\textwidth,trim=0.0cm 3.5cm 9.5cm 0.0cm,clip
    ]{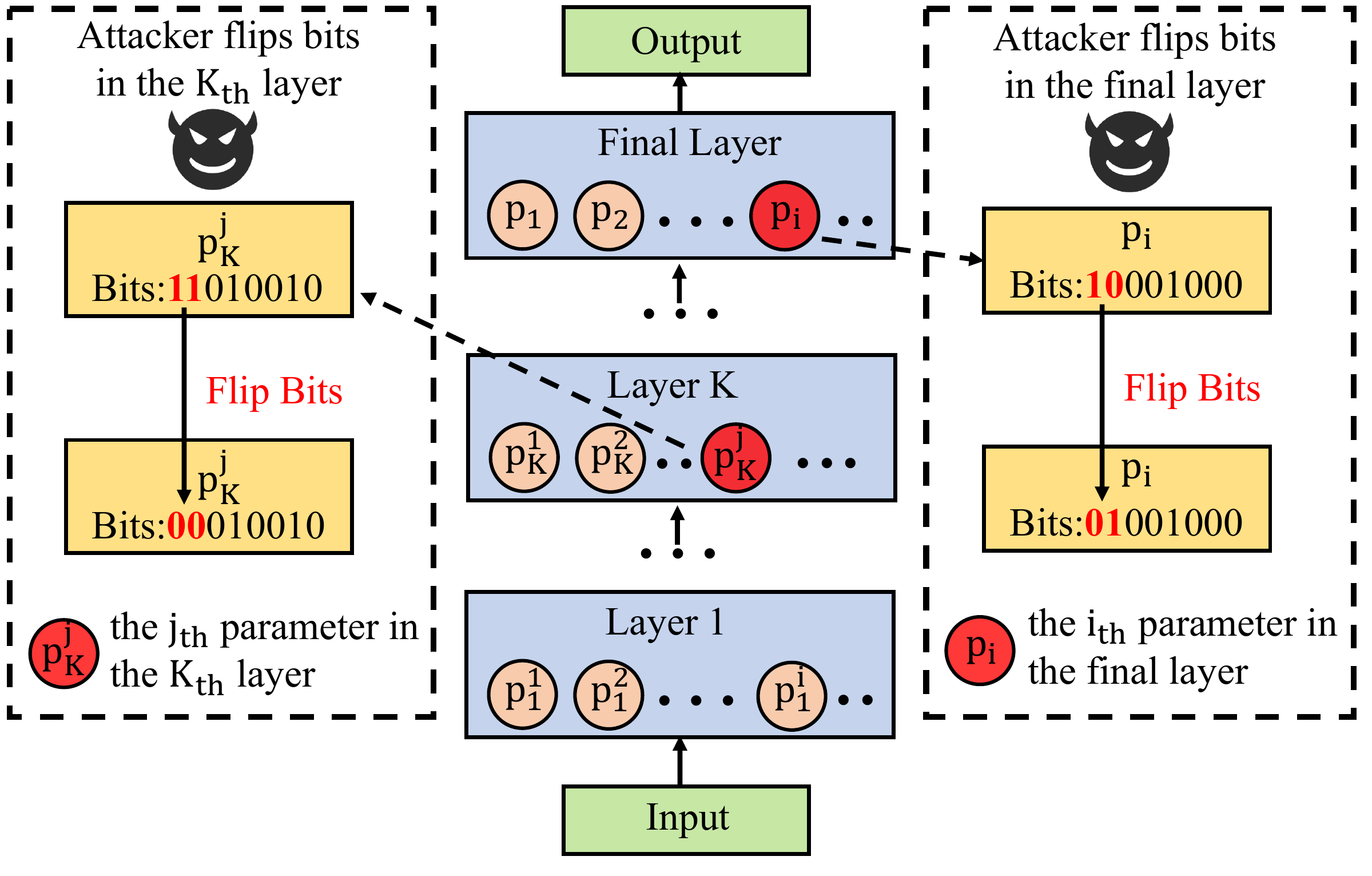} 
    %%%\vspace{-0.2cm}
    \vspace{-5ex}
    \caption{An example of flipping bits in the $K_{th}$ or final layer.}
    \label{fig:method_fliplast}
    \vspace{-2ex}
\end{figure}

First, there are BFAs that only flip bits in the final layer since the parameters of the final layer are directly related to the prediction results. 
It is straightforward to deduce that a multi-exit mechanism can thwart the basic BFAs that flip bits only in the final layer (as shown in Figure~\ref{fig:method_fliplast}). 
(1) Using a multi-exit DNN structure such as the SDN can interfere with the adversarial perturbations carried by the samples (triggers generated only from the vanilla model). 
Malicious samples may exit early to stop inference at an arbitrary hidden layer which generates different predictions compared with the inference on the target vanilla model. 
(2) By forcing most samples to exit early, the flipped bits at the final layer will be probably ignored during inference to achieve the defense goals. 

Second, we consider the existing more sophisticated BFAs that are not targeting only the final layer. 
For instance, the adversary may use an optimal way to locate the critical bits in the hidden layers (e.g. flipping bits in the $K_{th}$ layer in Figure~\ref{fig:method_fliplast}). 
Directly using the SDN structure cannot provide protection when the critical bits are flipped in the shallow layers. 
Moreover, in a white-box scenario, the adversary can observe the exit distribution of samples to locate critical exits for performing attacks in the corresponding critical layers (e.g. 78\% samples will exit in the last five layers of a VGG-based SDN model). 
Thus, we propose a Dynamic-Exit SDN (\desdn) mechanism that randomizes the exit for each inference. 
This \desdn can mitigate the case when the adversary flips bits in shallow layers since the sample exits the model in a random layer, which has a low probability of containing the flipped bits. Moreover, \desdn can push the exit distribution to a uniform one (see our experiments in~\figurename~\ref{fig:distribution} (a)) such that there are no critical exits for the adversary to consider.

Third, we further consider the most powerful adaptive attack. 
With the knowledge of all the details of \sysname, the adversary may include all exits to optimize his critical bit search. 
Although this will increase the attack cost (more bits to flip), it is possible to flip bits, particularly targeting at \sysname to achieve the attack regardless of where the sample exits. 
Our insight is to further design a robust training (\rob) strategy to find the critical and vulnerable bits for clean samples' inference process and simulate the influence when they are flipped. 
Note we only perform \rob on ICs without touching the target vanilla model to guarantee the non-intrusive defense requirement. 
This can improve the robustness of ICs to mitigate the significant convolutional output change when certain bits are flipped. 
Therefore, the difficulty of performing bit-flip attacks will be further increased.
Below we detail two core components (\desdn and \rob) of \sysname and analyze its security against various attacks.

\subsection{Dynamic-Exit SDN (\desdn)}
\label{DESDN}
As the first component of \sysname, \desdn consists of two steps: converting a model $M$ to an SDN model $\hat{M}$ (offline), and performing a random exit strategy during inference (online). 

\subsubsection{Stage 1: Constructing SDN Model}

\begin{figure}[!h]
    \centering
 \includegraphics[width=.45\textwidth,trim=0.0cm 7.6cm 0.2cm 0.3cm,clip]{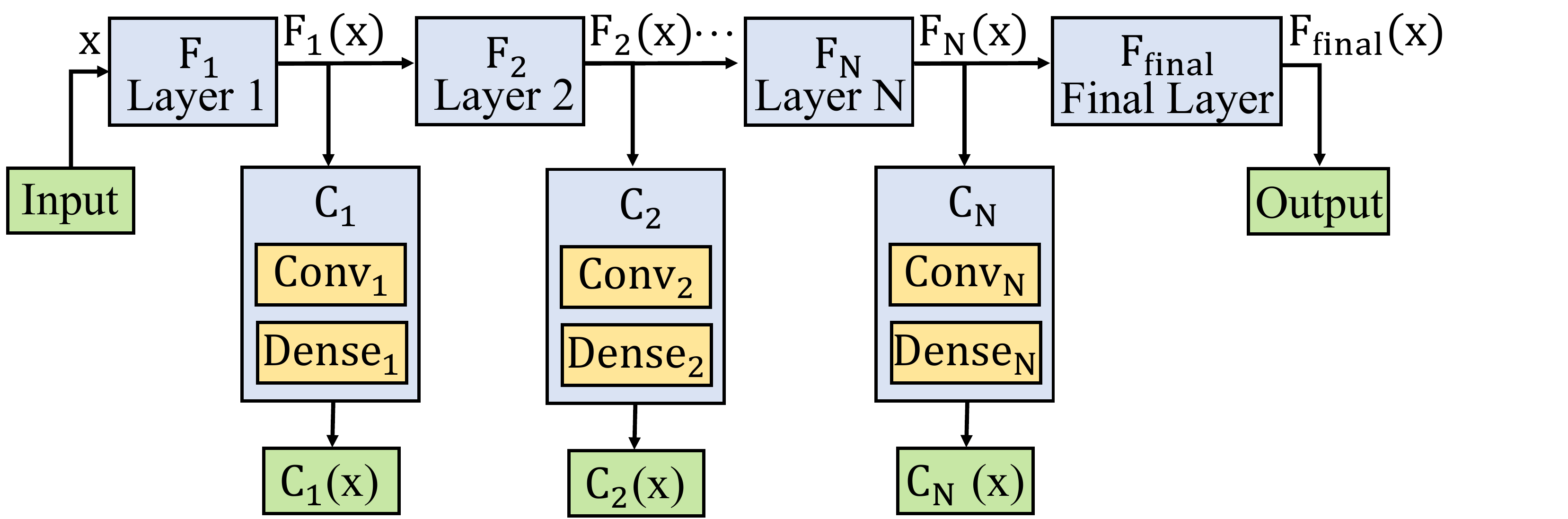} 
    \vspace{-2ex}
    \caption{An example of the model attached by ICs. $C_{i}$ denotes the $i$-th IC attached by us. During inference, each IC could make a prediction. For example, $C_{i}(x)$ is the prediction of $C_{i}$.}
    \label{fig:method_ic_fig}
    \vspace{-2ex}
\end{figure}

We adopt the technique in prior work~\cite{kaya2019shallow} to build the SDN model $\hat{M}$ from $M$, which shows negligible accuracy degradation during conversion. 
Specifically, we assume $M$ consists of $N$ internal layers $F_{i} $, $(1\leq i \leq N)$, and ends with the final layer $F_{final}$. For an inference sample, $M$ performs the classification as $M(x)=F_{final}(F_{N}(, ...F_{1}(x)))$. For simplicity, we denote the output of the $i$-th internal layer as $F_{i}(x)$, and the output of the final layer $M(x)$ as $F_{final}(x)$. Then our goal is to train an IC ($C_i$) for each internal layer $i$, as shown in  Figure~\ref{fig:method_ic_fig}. Then $C_i$ is attached to layer $i$ and makes the prediction $C_{i}(F_{i}(x))$, which is simplified as $C_{i}(x)$. To restrict the size of the IC, each $C_i$ only contains one convolutional layer and one dense layer. Such a simple structure makes it efficient to learn the parameters while maintaining high classification accuracy. This design is general and can be applied to different models. 

During construction, the defender freezes the parameters of the original model $M$ but just trains the ICs. Note, this training process is much more efficient than training a complete model from scratch. For instance, training ICs for a vanilla model is $3.2 \sim 8.2\times$ faster than training this model.
% Besides, IC training cost depends on datasets. 
% For example, the training cost on CIFAR-10 and ResNet32 is $12.2\%$ of training the original model, which indicates $8.2\times$ faster than before.
% In contrast, the training cost on Tiny-ImageNet and VGG16 is $3.2\times$ faster than before.
Further, there is a trade-off between IC training cost and model accuracy. For instance, if we allow a tiny accuracy drop (e.g. $2\%$ like previous work~\cite{DBLP:conf/cvpr/HeRLCF20}), IC training cost will be less than $10\%$ of the original model training and can be negligible.
We leave further reduction of training cost as our future work.
% (see Appendix~\ref{appendix-trainingcost} for thorough comparisons).

\subsubsection{Stage 2: Randomizing Exits During Inference} 
\label{Section-stage2}
After attaching the trained ICs to $M$, the constructed SDN model $\hat{M}$ allows early exit. 
%In the original SDN protocol~\cite{kaya2019shallow}, 
A threshold  $\tau$ is then introduced to judge whether the inference should exit at each internal layer $F_i$. Specifically, for a given sample $x$, when the inference process reaches the $i$-th layer, we compute the confidence score of the corresponding IC $max(C_{i}(x))$. If this score is larger than $\tau$, then the process will exit from $C_{i}$ with the corresponding output without going into deeper layers. This deterministic exit mechanism can thwart the basic BFAs, but may still be vulnerable to adaptive attacks.

To further secure the inference computation, we design a dynamic exit strategy. 
Particularly for each query sample $x$, among all the ICs $C=\{C_{0}, C_{1}, ..., C_{N}, F_{final}\}$, we randomly select a set of $q$ candidate ICs, denoted as $\hat{C}$. Then we perform the early exit within these candidate ICs based on their confidence scores: we find the first IC $C_i$ in $\hat{C}$ whose confidence score $max(C_{i}(x))$ is larger than the threshold $\tau_i$. Then this layer is selected as the early exit for this inference sample. If none of the candidate ICs can satisfy the early exit criteria, we will choose the final layer in $\hat{C}$ as the exit for prediction. 
There exists a trade-off between model accuracy and security, determined by the hyper-parameter $q$. Specifically, a smaller $q$ can make the selected exit more random with larger entropy. However, it also increases the probability that these $q$ ICs cannot meet the early exit criteria, and the prediction has lower confidence. In Section \ref{sec:evaluation} we will show that we can find the appropriate value of $q$ that brings high stochasticity to the exit selection, with negligible impact on the model accuracy.

\subsection{Robust Training on ICs}
\label{robust_train}
We then introduce the second core component of \sysname: \rob, which further enhances the defense effectiveness against BFAs. 
In particular, when the adversary flips bits in the $i$-th layer of $M$, it could still possibly affect any ICs after this layer with a certain probability, as the layer output $F_i(x)$ passes to these ICs and affects their predictions. 
To reduce such impacts, we propose \rob to improve the robustness of the attached ICs. 
Note here the \rob is only performed on ICs so there is no modification on the target model. 

\begin{figure*}[!h]
    \centering
 \includegraphics[width=.85\textwidth,trim=0.1cm 4.2cm 0.0cm 0.3cm,clip
    ]{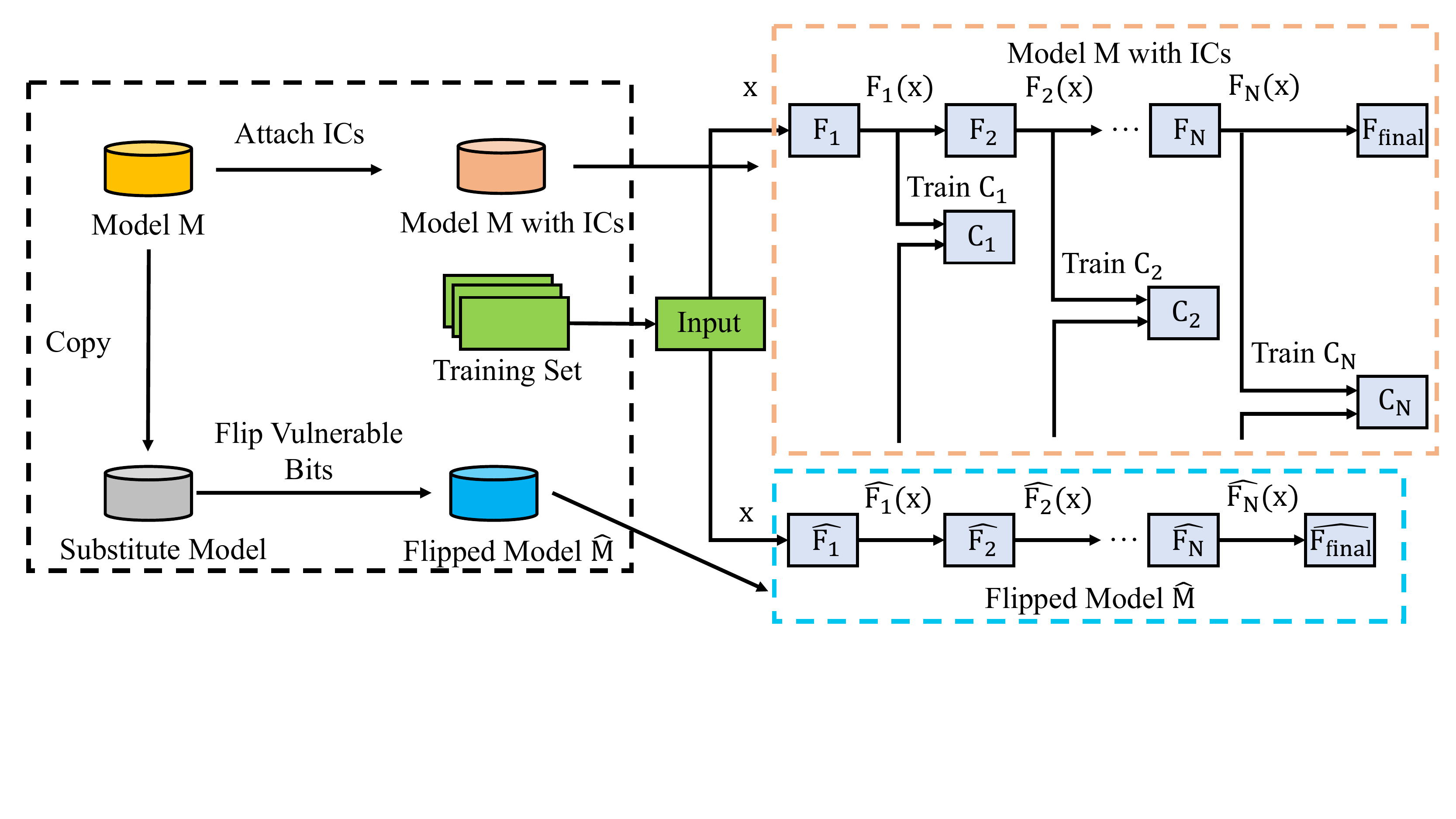} 
    \vspace{-2ex}
    \caption{Illustration of \rob. We first construct the flipped model $\hat{M}$, and then synthesize training samples for \rob. In particular, we use $\hat{M}$ to generate special training data, e.g., $\hat{F}_{1}(x)$. 
    Then, we apply these special training data, together with the original training data to train ICs. For example, we use ${F}_{1}(x)$ and ${\hat{F}}_{1}(x)$ to train $C_1$.}
    \label{fig:method_robust}
    \vspace{-3ex}
\end{figure*}

The key insight of \rob is to help ICs adapt to the cases when critical bits are flipped in $M$. 
In general, without prior knowledge of the BFAs, we construct a bit-flipped model by considering only the benign samples' inference process to simulate the target compromised model. 
Then, we craft new samples based on this bit-flipped model for \rob. 
These training samples simulate the outputs of flipped layers, and ICs will learn such data to correct the prediction from the adversarial scenarios.
Figure~\ref{fig:method_robust} illustrates the basic idea of this approach, which consists of two steps. 

The first step is to construct the flipped model. For defense generality, we assume the defender does not know the exact attack methods to mitigate. In this case, to make the flipped model closer to the real-world victim model, we design a vulnerable-protection algorithm (VPA) to figure out vulnerable bits that could be potentially flipped. 
Our VPA aims to find bits that are critical to the model decisions.
Such bits might significantly affect the prediction results and are vulnerable to being flipped by the adversary.
We note that existing attack methods~\cite{DBLP:conf/cvpr/RakinHF20,DBLP:conf/iclr/BaiWZL0X21,DBLP:conf/iccv/ChenFZK21,DBLP:conf/iccv/RakinHF19,yao2020deephammer} treat gradients w.r.t bits as a key component to select flipped bits.
Indeed, the gradients of the model output w.r.t bits reflect the importance of bits in model decisions.
Inspired by this, the basic idea of our VPA is to select critical bits according to the gradients of inference loss $\mathcal{L}_{inf}$  w.r.t bits, where $\mathcal{L}_{inf}$ is defined as follows:
\begin{equation}
\label{eq:infer_loss}
\begin{aligned}
    \mathcal{L}_{inf}&=\mathcal{L}_{ce}(F_{final}(x);l),
\end{aligned}
\end{equation} 
where $\mathcal{L}_{ce}$ is the cross-entropy loss, $l$ is the ground-truth label of the input sample $x$. Below we describe VPA in detail.

In particular, given a target model $M$, we denote all bits in $M$ as $B$. We first establish a substitute model which is exactly the same as $M$.
In each iteration of VPA,
(1) we follow previous work~\cite{DBLP:conf/iccv/RakinHF19} to calculate the gradients of $\mathcal{L}_{inf}$ w.r.t. each bit $b$ ($b \in B$), denoted as $\nabla_b{}\mathcal{L}_{inf}$.
(2) Then, we descendingly rank the vulnerability of bits by the absolute value of their $\nabla_b{}\mathcal{L}_{inf}$, and select the bits with the top-$k$ gradients.
(3) We treat these bits as vulnerable bits, and flip them.
We iterate the above process until the maximum iteration budget $N_{vpa}$ exhausts to get the flipped model.
Note that in each iteration, VPA must recalculate the gradients of each bit: as the model $M$ is dynamically changed due to the flipped bits, old gradients could not reflect the importance of bits in the newly changed model decisions. After this step, we can get the flipped model $\hat{M}$, which consists of $N$ internal layers $\hat{F}_{i}$, $(1\leq i \leq N)$.

The second step is to synthesize training samples for \rob. We freeze the original weights and only train weights in ICs.
Given an input sample $x$ from the original training set, we denote the output of the layer $i$ from the flipped model $\hat{M}$ as $\hat{F}_{i}(x)$. So $\hat{F}_{i}(x)$ will serve as the special training sample for the IC attached to layer $i$. It can help the IC better adapt to the attack scenarios in advance, so as to improve the robustness of the IC. During training, except for $\hat{F}_{i}(x)$, we also need to apply original training data ${F}_{i}(x)$ provided by $M$, so as to ensure the accuracy of ICs.

\subsection{Security Analysis} 
\label{defense_ana}

After introducing the design of \sysname,
we give a comprehensive qualitative analysis of the resilience of this methodology against different types of BFAs. The first two are basic attacks from existing works, while the last one depicts a possible adaptive strategy targeting our new defense mechanism. 

\noindent\textbf{Attack I}: the adversary only flips the bits in the final layer of the model. This strategy is adopted in TBT~\cite{DBLP:conf/cvpr/RakinHF20} and TA-LBF~\cite{DBLP:conf/iclr/BaiWZL0X21}. The basic SDN model can defeat these two attacks. 

Formally, the TBT attack defines a target class $t$ and a specific trigger $\Delta$ for activating the backdoor in the victim model. Given a clean input sample $x$ with the ground truth label $l$, the attacker aims to make the compromised model mispredict $x+\Delta$ as $t$. Identification of the critical bits can be modeled as an optimization problem, which aims to minimize the following loss function:
\begin{equation}
\label{eq:TBT_ori_loss}
\begin{aligned}
    \mathcal{L}_{TBT}=\mathcal{L}_{ce}(F_{final}(x);l)+\mathcal{L}_{ce}(F_{final}(x+\Delta);t),
\end{aligned}
\end{equation} 
where $\mathcal{L}_{ce}$ is the Cross-Entropy loss function. We observe that the loss function only considers the final layer. As a result, the basic SDN model is able to thwart this attack, as most inference samples will exit earlier before the final layer $F_{final}$, and not be affected by the flipped bits. 

Similarly, TA-LBF is a sample-wise attack, which aims to cause the misclassification of a specific sample $x$ from its ground truth label $l$ to the target label $t$. This can also be formulated as an optimization problem:
\begin{equation}
\label{eq:sam_ori_loss}
\begin{aligned}
    \mathcal{L}_{TA-LBF}= \mathcal{L}_1(x;l;t;F_{final}) + \lambda\mathcal{L}_2(x;l;t;F_{final}),
\end{aligned}
\end{equation} 
where $\lambda$ is a hyperparameter, $\mathcal{L}_{1}$ and $\mathcal{L}_{2}$ are two specific loss functions that ensure the attack effectiveness and stealthiness, respectively. We observe that this loss function is also only related to the final layer $F_{final}$. With an SDN, the sample $x$ has a high chance to exit the model earlier than the final layer, and will not be affected by the flipped bits in $F_{final}$.  

\noindent\textbf{Attack II}: the adversary flips bits in arbitrary layers based on a more sophisticated search method, as exploited in ProFlip~\cite{DBLP:conf/iccv/ChenFZK21}. Specifically, the adversary selects a trigger $\Delta$. Given a clean input sample $x$ and the target class $t$,
% the target class $t$  that $x+\Delta$ is aimed to be mispredicted to, 
the adversary aims to search for critical bits in all the layers to inject the backdoor. This process is also modeled as an optimization problem with the following loss function:
\begin{equation}
\label{eq:attack2_loss}
\begin{aligned}
    \mathcal{L}_{ProFlip}=\mathcal{L}_{ce}(F_{final}(x+\Delta);t)-\mathcal{N}(x+\Delta),
    % \mathcal{L}=\mathcal{L}_{mse}(\mathcal{N}(x+\Delta);t)+\mathcal{L}_{ce}(F_{final}(x);l),
\end{aligned}
\end{equation} 
where $\mathcal{N}$ denotes the salient neurons, determined by the target label through conducting the Jacobian Saliency Map Attack (JSMA)~\cite{DBLP:conf/eurosp/PapernotMJFCS16}.
Particularly, the adversary calculates the gradients of the model inference $F_{final}(x)$ w.r.t. each neuron output, and
then selects neurons with top gradient values as $\mathcal{N}$.

We observe that the joint impact of all the searched bits can only affect the output of $F_{final}$. Samples that exit earlier from the ICs might still give the correct prediction results.

\noindent\textbf{Attack III}: 
\label{sec:adap}
%%%%%%%%%%%%%%%%%%%%
based on the above analysis, we conclude that the basic BFAs that flip bits in certain layers cannot break the basic SDN. So we further assume a stronger adversary, who knows our \sysname mechanism and aims to design an adaptive strategy to break it. He may design a loss function $\mathcal{L}^*$ that considers all the ICs to optimize. However, breaking our defense is still difficult, as explained below:

(1) For adaptive attacks based on Attack I (i.e., adaptive TBT and TA-LBF), considering all the ICs could identify the vulnerable bits that can affect each exit. However, this could significantly increase the attack cost (i.e., the number of bits to flip). In particular, to design an adaptive TBT, we consider the following loss function, which includes the ICs as well:
\begin{equation}
\label{eq:TBT_adap_loss}
\begin{aligned}
    \mathcal{L}^*_{TBT} = \mathcal{L}_{TBT} +\sum_{i=1}^{N}\mathcal{L}_{ce}(C_{i}(x);l)+\mathcal{L}_{ce}(C_{i}(x+\Delta);t).
\end{aligned}
\end{equation} 
In this case, the adversary flips bits in the final layer of each IC as well as the final layer of $M$. Recall that in the TBT algorithm in Section~\ref{background:Targeted}, the adversary sets a fixed number of candidate parameters in the final layer as $w_b$.
For each candidate parameter, the adversary figures out several bits to flip.
This implies that the number of flipped bits is positively correlated with $w_b$.
In adaptive scenarios, since each IC needs to be attacked, the adversary needs to modify $w_b$ parameters in the final layer of each IC. This results in the scale of $n\times w_{b}$ parameters for compromising, which is a huge cost for the adversary, especially when $n$ is large.

The adaptive TA-LBF can also be designed in a similar way by including the ICs in the loss function as follows:
\begin{equation}
\label{eq:sam_adap_loss}
\begin{aligned}
        \mathcal{L}^*_{TA-LBF} = \mathcal{L}_{TA-LBF} +\sum_{i=1}^{N}\mathcal{L}_{1}(x;l;t;C_{i})+\mathcal{L}_{2}(x;l;t;C_{i}).
\end{aligned}
\end{equation} 
Increasing the number of candidate layers also results in a larger number of bits to flip, since the adversary needs to ensure all the exits are affected.

%\noindent\textbf{Adaptive ProFlip:}
(2) For adaptive attacks based on Attack II (i.e., adaptive ProFlip), although the adversary considers all the ICs, he might still fail to attack each exit. Samples exiting from the unattacked ICs are not affected, and the attacks are mitigated.
In particular,
for the adaptive ProFlip attack, the adversary optimizes the following loss function:
\begin{equation}
\label{eq:pro_adap_loss}
\begin{aligned}
    \mathcal{L}^*_{ProFlip}&=\mathcal{L}_{ce}(F_{final}(x);t)+\sum_{i=1}^{N}\mathcal{L}_{ce}(C_i(x);t)-\mathcal{\widetilde{N}}(x+\Delta).
\end{aligned}
\end{equation} 
Different from ${\mathcal{N}}$, 
the adversary considers all ICs as well as the final layer, i.e., $\sum_{i=1}^{N}C_i(x)+F_{final}(x)$.
Particularly,
the adversary adopts JSMA to calculate the gradients of $\sum_{i=1}^{N}C_i(x)+F_{final}(x)$ w.r.t each neuron outputs, then select neurons with top gradients, denoted as $\widetilde{\mathcal{N}}$.

The adversary selects the optimal parameter to modify through optimizing Eq.~\ref{eq:pro_adap_loss}, and the optimal parameter might locate in any layer. 
Here, the determination of the optimal parameter is highly dependent on $x$ given $t$, $\Delta$, and the target model.
For example, on CIFAR-100 and VGG16, we repeat the ProFlip attack 100 times, and randomly select 256 different input samples $x$ for each time.
For each independent attack, we observe the optimal parameter could be determined in different layers of $\hat{M}$. 
We assume the optimal parameter is located in the $i$-th layer.
In this case, all ICs attached before the $i$-th layer are not destroyed by the adversary.
Thus the samples exiting from these ICs are not affected.
Besides, for the ICs attached after the $i$-th layer, our \rob could reduce the impacts brought by the flipped bits.  
Another case is that the optimal parameter is located in an IC. Only the IC is attacked by the adversary, and all other ICs are not affected.

\section{Evaluation}
\label{sec:evaluation}\vspace{-1ex}
We evaluate \sysname on different targeted BFAs: two backdoor attacks (TBT and TA-LBF) and one sample-wise attack (ProFlip). 
We also consider many adaptive attacks. 
All experiments are conducted on a machine with an Intel Xeon Gold 6154 CPU and 8 NVIDIA Tesla V100 GPUs. \vspace{-1ex}

\subsection{Experimental Setup}
~\label{sec5-1}\vspace{-3ex}

\noindent\textbf{Datasets and models.} We conduct our experiments on four widely-used datasets with two DNN structures: VGG16~\cite{DBLP:journals/corr/SimonyanZ14a} and ResNet32~\cite{DBLP:conf/cvpr/HeZRS16}.

\begin{packeditemize}

\item CIFAR-10~\cite{krizhevsky2009learning}: 
This dataset contains 50,000 training images and 10,000 test images. 
Each image has a size of $32\times32\times3$ and belongs to one of 10 classes.

\item CIFAR-100~\cite{krizhevsky2009learning}: 
This has the same number of training and testing images as CIFAR-10. Each image also has the same size as CIFAR-10, but belongs to one of 100 classes. 

\item STL-10~\cite{DBLP:journals/jmlr/CoatesNL11}: 
This dataset contains 50,00 training images and 8,000 test images. 
We note that it also contains 100,000 unlabeled images.
Each image has a size of $96\times96\times3$ and belongs to one of 10 classes.
%We train target our models for STL-10 with VGG16~\cite{DBLP:journals/corr/SimonyanZ14a} and ResNet32~\cite{DBLP:conf/cvpr/HeZRS16}.

\item Tiny-ImageNet~\cite{tiny-img}: 
This dataset is a simplified version of ImageNet consisting of color images with a size of $64\times64\times3$ belonging to 200 classes. 
Each class has 500 training images and 50 testing images.

\end{packeditemize}

%\noindent\textbf{Dataset splitting.}
Inspired by~\cite{arp2022and}, we strictly separate the training and testing data without any overlap.
In particular, (1) CIFAR-10:  we follow~\cite{carlini2017towards,DBLP:conf/cvpr/HeRLCF20,DBLP:conf/issta/Wang0RYLLZ22,DBLP:conf/icml/MoonAS19} to select 50,000/10,000 images for training/testing. 
(2) CIFAR-100: we follow~\cite{DBLP:conf/nips/KhoslaTWSTIMLK20,DBLP:conf/nips/BelloFDCSLSZ21} to select 50,000/10,000 images for training/testing. 
(3) STL-10: we follow~\cite{DBLP:conf/nips/SwerskySA13,DBLP:conf/nips/Tian0PKSI20} to select 5,000/8,000 images for training/testing. 
(4) Tiny-ImageNet: we follow~\cite{DBLP:conf/issta/Wang0RYLLZ22,DBLP:conf/nips/Tian0PKSI20} to select 100,000/10,000 images for training/testing.

\noindent\textbf{Hyperparameters.}
% As mentioned in Section~\ref{DESDN}, 
As mentioned in Section~\ref{Section-stage2}, $\tau$ and $q$ affect the early-exit distribution and model accuracy. 
% We need to set the early-exit threshold $\tau$\footnote{We follow the initial SDN ~\cite{kaya2019shallow} to set the same $\tau$ for all the ICs given one target model. Tuning this hyperparameter has a tiny influence on our results.}, and the number $q$ of ICs randomly selected for each inference.
For generalization on unseen data and avoiding the selection of biased hyperparameters,
we tune hyperparameters on training data guided by two goals: (1) making early-exits uniformly distributed to prevent the attacker from targeting only those popular exits; (2) maintaining high ACC on benign samples.
Table~\ref{tab:hyperparameter} list the values of these hyperparameters.
Note that we tune these hyperparameters without considering any specific attacks. 
They are general and fixed to mitigate all attacks in our consideration. 
% We also evaluate the sensitivity of these hyperparameters in Appendix~\ref{appendix:hyper_sensitivity} and find the mitigation results are stable to the hyperparameters that meet the two goals. 
We also evaluate the sensitivity of these hyperparameters and find the mitigation results are stable to hyperparameters that meet the two goals (see  Appendix~\ref{appendix:hyper_sensitivity}).

\begin{table}[htbp]\vspace{-2ex}
\centering
\caption{Values of $\tau$ and $q$ on our datasets and models.}
%%\vspace{-3ex}
\label{tab:hyperparameter}
\resizebox{0.6\linewidth}{!}{
\begin{tabular}{cccc}
\Xhline{1pt}
\textbf{Dataset} & \textbf{Model} & \textbf{$\bm{\tau}$} & \textbf{$q$} \\\Xhline{1pt}
\multirow{2}{*}{CIFAR-10} & ResNet32 & 0.95 & 3 \\\cline{2-4}
 & VGG16 & 0.95 & 3 \\ \hline
\multirow{2}{*}{CIFAR-100} & ResNet32 & 0.90 & 5 \\\cline{2-4}
 & VGG16 & 0.80 & 3\\ \hline
 \multirow{2}{*}{STL-10} & ResNet32 & 0.95 & 5 \\\cline{2-4}
 & VGG16 & 0.95 & 3 \\ 
 \hline
 \multirow{2}{*}{Tiny-ImageNet} & ResNet32 & 0.90 & 3 \\\cline{2-4}
 & VGG16 & 0.95 & 4 \\ 
 \Xhline{1pt}
\end{tabular}\vspace{-4ex}
}
\end{table}

% \noindent\textbf{Hyperparameters.}

We also conduct \rob for the ICs.
The only randomness of Aegis is from the random selection of ICs in inference. We repeated each experiment 10 times and the ASR variance is below $2\%$ which does not affect our conclusion.

\noindent\textbf{Baselines.}
We compare \sysname with the state-of-the-art defense methods BIN ~\cite{DBLP:conf/cvpr/HeRLCF20} and RA-BNN~\cite{rakin2021ra}. 
We also compare \sysname with the basic SDN~\cite{kaya2019shallow} to demonstrate the effectiveness of \desdn and \rob mechanisms.
For a fair comparison, we slightly modify SDN to make sure \sysname uses the same structure and hyperparameters as SDN, except for the \desdn and \rob mechanisms. 
Besides, we compare \sysname with the baseline models (BASE) with no defense.

\subsection{Model Utility Evaluation}
\label{sec:acc}
A qualified defense method should preserve the utility, i.e. tiny model accuracy (ACC) drop. 
Table~\ref{table:acc} compares the impacts of different methods on ACC. 
We observe that \sysname slightly degrades ACC by approximately less than $2\%$, while BIN and RA-BNN have much more ACC degradation, i.e., roughly $2\sim12\%$ and $2\sim7\%$, respectively. 
This is due to BIN and RA-BNN adopting an aggressive binarization on weights (each parameter occupies only 1 bit) or activation function outputs which harm the model ACC. 
We also notice that SDN has comparable ACC with \sysname, which validates that the \desdn mechanisms in \sysname do not affect the model utility. 
In summary, \sysname could preserve the model utility.

\begin{table}[!htbp]\vspace{-3ex}
\centering
\caption{Model ACC influence evaluation. }
\label{table:acc}
%%\vspace{-3ex}
\resizebox{0.99\linewidth}{!}{
    \begin{tabular}{c|c|c|c|c|c|c} \Xhline{1pt}
    \multirow{2}*{\textbf{Dataset}} &
    \multirow{2}*{\textbf{Model}}  & \textbf{BASE} &\multicolumn{4}{c}{{}\textbf{$\Delta$~ACC~(\%)}} 
    % & \multicolumn{3}{c}{\textbf{VGG16}} 
    \\\cline{4-7}
    &    & \textbf{ACC (\%)}  & \textbf{BIN}& \textbf{RA-BNN} & \textbf{SDN} & \textbf{\sysname}  \\ \Xhline{1pt}
    \multirow{2}{*}{CIFAR-10}  & ResNet32
& 92.79 &-2.26& -1.71&-1.27&\textbf{-1.26}\\  \cline{2-7}
&VGG16&93.61&-1.26&-1.19&-1.72& \textbf{-0.67}\\ \cline{2-7}
 \hline
 \multirow{2}{*}{CIFAR-100}  & ResNet32
& 66.13 &-4.38&-2.47&-2.54&\textbf{-1.96}\\  \cline{2-7}
&VGG16&72.85 &-4.14&-2.08&-1.97& \textbf{-1.90}\\ \cline{2-7}
 \hline
\multirow{2}{*}{STL-10}  & ResNet32
& 74.80&-4.09&-3.85&-2.80&\textbf{-0.90}\\ 
 \cline{2-7}
&VGG16& 79.51&-1.41&-1.39&-1.35& \textbf{-1.02}\\ 
 \cline{2-7}
 \hline
\multirow{2}{*}{Tiny-ImageNet}  & ResNet32
&54.58 &-11.16&-6.31&-3.87&\textbf{-1.92}\\ 
 \cline{2-7}
&VGG16& 60.51&-4.18&-4.07&-0.39& \textbf{-0.28}\\ 
 \cline{2-7}

    \Xhline{1pt}
    \end{tabular}}\vspace{-2ex}
    
\end{table}

\subsection{Mitigating Targeted Attacks}
~\label{sec5-3}
We evaluate the defense effectiveness of \sysname against the state-of-the-art targeted attacks including two backdoors targeted BFAs (TBT~\cite{DBLP:conf/cvpr/RakinHF20} and ProFlip~\cite{DBLP:conf/iccv/ChenFZK21}) and one sample-wise targeted BFA (TA-LBF~\cite{DBLP:conf/iclr/BaiWZL0X21}). 
We reproduce these BFAs with their open-sourced code and the recommended parameters (Appendix~\ref{appendix-attacks}) and list the visual results (e.g. triggers and samples) in Appendix~\ref{appendix:vis_trigger}. 
The possible adaptive attacks based on these BFAs are evaluated in Section~\ref{exp:adap}. 

\noindent\textbf{Metrics.}
\label{compare_metrics}
While applying an attack to all defense methods, we compare the attack success rate (ASR) of each defense method.
%The lower the ASR, the better the defense method.
For fair comparisons, we make sure the given attack pays the same attack cost on each defense method: flipping the same number of bits.
We denote the number of flipped bits as $N_{b}$ and consider different $N_{b}$ with two steps. 

%MarkbyHan
First, we restrict the bit flipping number limit ($N_b$) to 50 for all attacks. 
Such a bit flipping number limit is set as 24 in the state-of-the-art Rowhammer attack on cloud platforms~\cite{jattke2022blacksmith}: it tests a batch of high-quality dual inline memory modules (DIMMs) and reveals that flipping 24 bits needs a significantly long time (several hours). 
We also confirm the feasibility of performing BFAs on physical systems (in Section~\ref{sec:feasibility}) to illustrate that flipping 50 bits is very difficult for attackers. 
Furthermore, we relax the restriction on $N_{b}$ (i.e. 500) and evaluate the ASR for comprehensiveness.

In the following, we first evaluate all defense methods under TBT and TA-LBF, as they all flip bits in the final layer of the target model.
We then evaluate all defense methods under ProFlip, which could flip bits in any layer of the target model.
% As we have mentioned above, we first restrict the $N_{b}$ to 50 for all attacks. 

\noindent\textbf{Mitigating TBT.}
\label{exp:no-tbt}
Table~\ref{tab:tbt-asr} shows the defense results against TBT.
\sysname can significantly decrease the ASR and outperform other methods. 
In most cases, \sysname can decrease the ASR to less than $20\%$, which is significantly lower than others.
In contrast, we note BIN and RA-BNN even perform worse than BASE in some cases. 
For example, on CIFAR-10 and VGG16, the ASR for BASE is $71.1\%$, which is lower than that of BIN ($90.4\%$) and RA-BNN ($82.9\%$).
This means the defenses designed for untargeted BFAs might make models even more vulnerable to targeted BFAs.

% \vspace{-3ex}
\begin{table}[!h]
\vspace{-3ex}
\centering
\caption{Evaluation results of ASR against TBT. }
\label{tab:tbt-asr}
%%\vspace{-3ex}
\resizebox{0.95\linewidth}{!}{
    \begin{tabular}{c|c|c|c|c|c|c} \Xhline{1pt}
    \multirow{2}*{\textbf{Dataset}} &
    \multirow{2}*{\textbf{Model}}  & \multicolumn{5}{c}{\textbf{ASR (\%)}} 
    % & \multicolumn{3}{c}{\textbf{VGG16}} 
    \\\cline{3-7}
    &
     & \textbf{BASE} & \textbf{BIN} & \textbf{RA-BNN}& \textbf{SDN} & \textbf{\sysname}  \\ \Xhline{1pt}
    \multirow{2}{*}{CIFAR-10}  & ResNet32
 &70.7&94.8&74.5&\textbf{16.3}& {19.9}\\  \cline{2-7}
&VGG16 &71.1&90.4&82.9&42.6& \textbf{36.0}\\ \cline{2-7}
 \hline
 \multirow{2}{*}{CIFAR-100}  & ResNet32
 &95.8&99.8&25.5&20.5& \textbf{10.8}\\  \cline{2-7}
&VGG16 & 65.9&58.4&47.4&53.8&\textbf{10.6}\\ \cline{2-7}
 \hline
\multirow{2}{*}{STL-10}  & ResNet32
&100.0&72.5&29.4&47.1& \textbf{13.0}\\ 
 \cline{2-7}
&VGG16& 64.1&99.7&88.0&\textbf{9.0}& {10.5}\\ 
 \cline{2-7}
 \hline
\multirow{2}{*}{Tiny-ImageNet}  & ResNet32
 &100.0&63.3&31.4&65.8&\textbf{ 27.9}\\ 
 \cline{2-7}
&VGG16& 69.7&72.3&40.2&48.9& \textbf{10.1}\\ 
 \cline{2-7}

    \Xhline{1pt}
    \end{tabular}}\vspace{-1ex}
\end{table}

% The ASR for BASE is $64.1\%\sim100.0\%$, and the ASR for BIN is higher than $60\%$ in most cases. 
% This proves that as an effective defense scheme against untargeted BFAs, BIN will make models even more vulnerable to targeted BFAs. 
% Then, taking STL-10 and ResNet32 as an example, \sysname reduces the ASR to $13.0\%$. 
% However, the ASR for BASE, BIN, RA-BNN, and SDN is $100.0\%$, $72.5\%$, $29.4\%$, and $47.1\%$, respectively. 
% This evaluation proves the effectiveness of \sysname against TBT attack.

\noindent\textbf{Mitigating TA-LBF. }
Table~\ref{tab:sample-asr} shows the ASR on all datasets and models against TA-LBF.
Overall, \sysname could effectively mitigate TA-LBF and outperform other methods. 
In most cases, \sysname can limit ASR below $10.0\%$, which is much smaller than others. 
For example, on STL-10 and ResNet32,  \sysname limits the ASR to $9.6\%$ while the ASR for BASE, BIN, RA-BNN, and SDN is $100.0\%$, $100.0\%$, $100.0\%$ and $47.7\%$, respectively. 
We find the highest ASR for \sysname is $20.1\%$ on Tiny-ImageNet and ResNet32 which is still much smaller than the ASR for other cases.

\begin{table}[!h]\vspace{-2ex}
\centering
\caption{Evaluation results of ASR against TA-LBF. }
\label{tab:sample-asr}
%%\vspace{-3ex}
\resizebox{0.95\linewidth}{!}{
    \begin{tabular}{c|c|c|c|c|c|c} \Xhline{1pt}
    \multirow{2}*{\textbf{Dataset}} &
    \multirow{2}*{\textbf{Model}}  & \multicolumn{4}{c}{\textbf{ASR (\%)}} 
    % & \multicolumn{3}{c}{\textbf{VGG16}} 
    \\\cline{3-7}
    &
     & \textbf{BASE} & \textbf{BIN} & \textbf{RA-BNN}& \textbf{SDN}& \textbf{\sysname}   \\ \Xhline{1pt}
    \multirow{2}{*}{CIFAR-10}  & ResNet32
 &100.0&100.0&100.0 &\textbf{3.5}& {6.3}\\  \cline{2-7}
&VGG16 &57.6&100.0&100.0 &1.1& \textbf{0.3}\\ \cline{2-7}
 \hline
 \multirow{2}{*}{CIFAR-100}  & ResNet32
 &100.0&100.0&100.0&38.0& \textbf{16.4}\\  \cline{2-7}
&VGG16 & 56.4&100.0&100.0&19.4&\textbf{4.4}\\ \cline{2-7}
 \hline
\multirow{2}{*}{STL-10}  & ResNet32
&100.0&100.0&100.0&47.7& \textbf{9.6}\\ 
 \cline{2-7}
&VGG16& 81.4&99.7&98.7&\textbf{0.3}& {2.0}\\ 
 \cline{2-7}
 \hline
\multirow{2}{*}{Tiny-ImageNet}  & ResNet32
 &100.0&100.0&100.0&71.1&\textbf{ 20.1}\\ 
 \cline{2-7}
&VGG16 &51.8&98.1&90.7&27.2& \textbf{ 17.3}\\ 
 \cline{2-7}

    \Xhline{1pt}
    \end{tabular}}\vspace{-2ex}
\end{table}

\noindent\textbf{Summary for the evaluation of TBT and TA-LBF.} 
We summarize the defense evaluation against TBT and TA-LBF as follows. 
These two attacks only flip bits in the final layer, which could not affect ICs.
\sysname can keep the model ACC well and let many input samples early-exit from ICs to effectively mitigate TBT and TA-LBF. 
We notice that SDN also performs better than BASE, BIN, and RA-BNN in most cases. 
However, SDN uses a static multi-exit mechanism, making it less effective than \sysname for adaptive attacks (Section~\ref{exp:adap}).

\begin{table}[!htbp]\vspace{-2ex}
\centering
\caption{Evaluation results of ASR against ProFlip. }
\label{tab:pro-asr}
%%\vspace{-3ex}
\resizebox{0.95\linewidth}{!}{
    \begin{tabular}{c|c|c|c|c|c|c} \Xhline{1pt}
    \multirow{2}*{\textbf{Dataset}} &
    \multirow{2}*{\textbf{Model}}  & \multicolumn{5}{c}{\textbf{ASR (\%)}} 
    % & \multicolumn{3}{c}{\textbf{VGG16}} 
    \\\cline{3-7}
    &
     & \textbf{BASE} & \textbf{BIN} & \textbf{RA-BNN}& \textbf{SDN}& \textbf{\sysname}   \\ \Xhline{1pt}
    \multirow{2}{*}{CIFAR-10}  & ResNet32
 &96.9&99.4&90.6&47.3& \textbf{19.8}\\  \cline{2-7}
&VGG16 &88.2&78.6&84.6&70.5& \textbf{28.9}\\ \cline{2-7}
 \hline
 \multirow{2}{*}{CIFAR-100}  & ResNet32
 &89.8&100.0&82.9&58.3& \textbf{19.2}\\  \cline{2-7}
&VGG16 & 80.0&80.4&76.5&64.9&\textbf{20.3}\\ \cline{2-7}
 \hline
\multirow{2}{*}{STL-10}  & ResNet32
&77.4&52.4&91.2&58.1& \textbf{33.9}\\ 
 \cline{2-7}
&VGG16& 87.2&96.0&90.3&19.9& \textbf{18.7}\\ 
 \cline{2-7}
 \hline
\multirow{2}{*}{Tiny-ImageNet}  & ResNet32
 &99.1&82.5&80.4&75.0&\textbf{ 20.1}\\ 
 \cline{2-7}
&VGG16 &88.2&44.1&39.2&26.8& \textbf{ 15.6}\\ 
 \cline{2-7}

    \Xhline{1pt}
    \end{tabular}}\vspace{-2ex}
\end{table}

\noindent\textbf{Mitigating ProFlip.} 
%We further evaluate \sysname against ProFlip.
Different from TBT and TA-LBF, ProFlip does not restrict the layer but uses an optimization method to flip critical bits in arbitrary layers. 
Tables~\ref{tab:pro-asr} shows the results in which \sysname can limit the ASR below $30\%$ in most cases.
On the contrary, the ASR for BASE, BIN, and RA-BNN is higher than $70\%$ in most cases.
For example, on CIFAR-10 dataset with VGG16, \sysname limits the ASR to $28.9\%$. However, the adversary acquires $88.2\%$, $78.6\%$, and $84.6\%$ ASR for BASE, BIN, and RA-BNN. 
\sysname makes that the flipped bits could not effectively affect samples that exit earlier. 
This is also why SDN has comparable defense performance in some cases. 
%However, SDN is much inferior to \sysname when considering adaptive attacks (Section~\ref{exp:adap}).

\begin{figure*}[!ht]
    \centering
    \includegraphics[width=.95\textwidth,
    trim=0.cm 9.8cm 0.cm 0.1cm,clip]{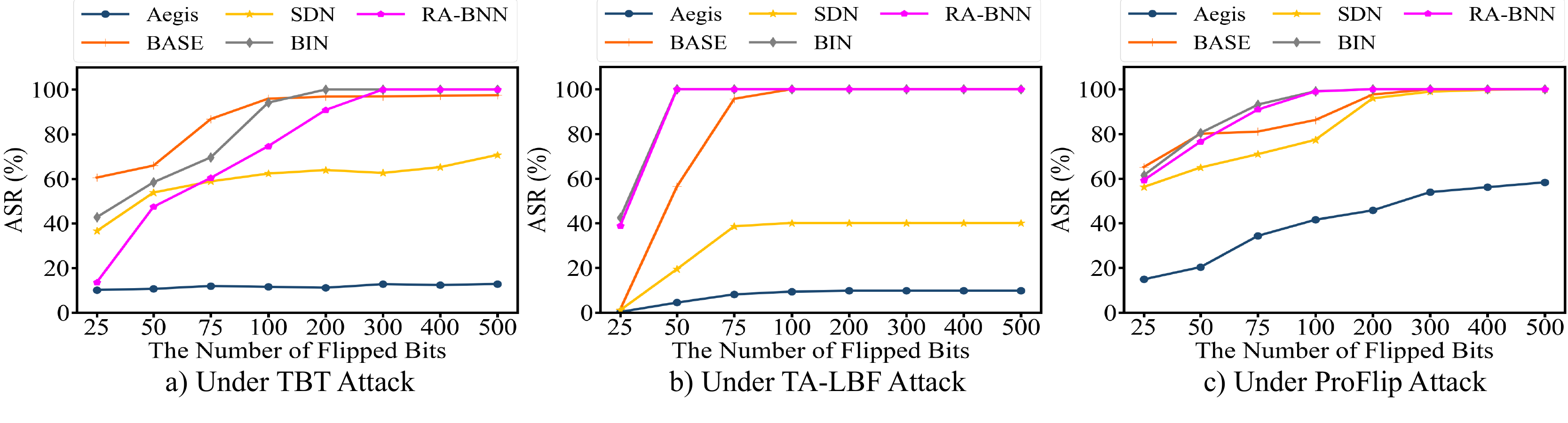} 
    \vspace{-4ex}
    \caption{Comparison between \sysname and other defense methods on CIFAR-100 with VGG16 when more bits are flipped. \sysname performs better than other methods: the ASR for \sysname is much lower than others under different numbers of flipped bits.}	
    \label{fig:asr_n_nonadap}
    \vspace{-2ex}
\end{figure*}

\noindent\textbf{Evaluation with more bits flipped.} 
Here we do not restrict $N_{b}$ to 50 but consider larger $N_{b}$ to compare \sysname with other methods. 
We take CIFAR-100 and VGG16 as an example to evaluate when $N_{b}$ increases to up to 500 for all defense methods. 
Evaluation results are shown in~\figurename~\ref{fig:asr_n_nonadap}.
Compared with existing defense methods, we observe the ASR for \sysname is still limited at a low level for all attacks. 
For TBT (\figurename~\ref{fig:asr_n_nonadap} (a)) and TA-LBF (\figurename~\ref{fig:asr_n_nonadap} (b)), even the attacker could flip as many as 500 bits, the ASR is still less than 13\%, which proves the defense effectiveness of our method. 
For ProFlip (\figurename~\ref{fig:asr_n_nonadap} (c)), ASR increases slowly with more bits flipped and \sysname can clearly outperform all baselines. 
Even if the attacker could flip as many as 500 bits, \sysname can restrict the ASR of ProFlip to 58.3\% while ASR for other cases is already 100\% with significantly fewer bits flipped. 

Note that when more bits are flipped, \sysname clearly outperforms SDN against all three attacks. 
Especially, considering ProFlip, the ASR achieves almost 100\% when about 200$\sim$300 bits are flipped. 
Thus, we claim that simply deploying a multi-exit DNN structure on a target vanilla model cannot sufficiently defeat these targeted BFAs even without considering their adaptive attacks. 
Also, \sysname includes two components, and the ablation study to evaluate their defense effectiveness respectively is given in Section~\ref{sec-abl-robust}. 

\iffalse
Note that, even if the attackers significantly increase $N_b$, \sysname still effectively mitigates all these attacks. 
Particularly, for TBT and TA-LBF, we restrict their ASR below $13.0\%$ under different $N_b$.
These two attacks only flip bits in the final layer of the victim models, thus the \desdn mechanism adopted by us could effectively mitigate them.
For ProFlip, we also restrict its ASR to a much lower level than that of other defense methods.
For example, until $N_{b}$ is 500, the ASR for our \sysname is $58.3\%$, while the ASR for other baselines is $100\%$.
For this model, ProFlip determines the optimal parameter in the $10_{th}$ layer and flip bits in this layer.  
Thus, the ICs before this layer are not affected by ProFlip.
% Besides, our robustness training contributes to the defense.
\fi

\subsection{Mitigating Adaptive Attacks}
\label{exp:adap}
Beyond basic adversaries, any effective defense should also be capable of withstanding any adaptive attackers who are aware of the existence and mechanism of the defense.  
We consider a sophisticated adversary who knows the detail of our defense mechanism and aims to design an adaptive strategy to break it. 
We consider crafting such advanced attacks from the three state-of-the-art attack methods (i.e., TBT, TA-LBF, and ProFlip).
%but try to adopt their attack strategy on \sysname. 
As analyzed in Section~\ref{sec:adap}, the adversary knows all details of \sysname and tries to attack all ICs to increase the ASR regardless of where the input samples exit. 
He can design new loss functions dedicated to \sysname by adding optimization terms for all ICs. 
Thus, the adaptive TBT and TA-LBF attacks will flip bits in the final layer of all ICs in addition to the final layer of the model. 
All other attack settings and models are the same as in the previous sections. 
%Note that this is the strongest adversary in our threat model since not only the attackers have the white-box access to the target models and all our defense schemes, but also the adaptive BFAs are dedicatedly modified and designed. 

In the following, we evaluate the effectiveness of \sysname against these adaptive attack scenarios. We still first set $N_b$ to 50 for all attacks. 
Then, we relax the restriction of $N_{b}$ and evaluate the ASR when the adversary could flip more bits.  

\noindent\textbf{Baselines.}
Since BIN is proven to make the victim model even more vulnerable to targeted BFAs, we do not consider it as a baseline defense and mainly compare with SDN.
Both \sysname and SDN have ICs attached to the hidden layers for an early-exit mechanism, enabling the adversary to design a similar dedicated adaptive attack against them.
%It is challenging for these two defense methods to defend against the adaptive attacks that consider all ICs. 
We also report the ASR of BASE to reflect the defense effectiveness. 

\noindent\textbf{Adaptive TBT.}
Table~\ref{tab:adap-tbt-asr} shows the ASR of each defense method. 
We observe that SDN can be defeated by the adaptive TBT attack.
By including all ICs of the SDN model to perform the adaptive TBT attack, the ASR of SDN is even higher than that of BASE in many cases. In contrast, our \sysname can restrict the ASR below 40\% in most cases. 
% For instance, on STL-10 and VGG16, the ASR of \sysname is $27.0\%$, while the ASR of SDN is $93.0\%$, which is much higher than the ASR of BASE  (i.e., $64.1\%$).
This is because the adaptive TBT attack will flip bits for critical parameters in ICs and the model's final layers to manipulate inference on all exits of the SDN model. 
However, \sysname still effectively mitigates this strategy due to the random exit mechanism.

\begin{table}[!htbp]\vspace{-3ex}
\centering
\caption{Evaluation results of ASR against adaptive TBT. }
\label{tab:adap-tbt-asr}
%%\vspace{-3ex}
\resizebox{0.75\linewidth}{!}{
    \begin{tabular}{c|c|c|c|c} \Xhline{1pt}
    \multirow{2}*{\textbf{Dataset}} &
    \multirow{2}*{\textbf{Model}}  & \multicolumn{3}{c}{\textbf{ASR (\%)}} 
    % & \multicolumn{3}{c}{\textbf{VGG16}} 
    \\\cline{3-5}
    & & \textbf{BASE}& \textbf{SDN} & \textbf{\sysname}  \\ \Xhline{1pt}
    \multirow{2}{*}{CIFAR-10}  & ResNet32
 &70.7&37.2& \textbf{31.1}\\  \cline{2-5}
&VGG16 &71.1&86.5& \textbf{58.1}\\ \cline{2-5}
 \hline
 \multirow{2}{*}{CIFAR-100}  & ResNet32
&95.8&79.3& \textbf{49.7} \\  \cline{2-5}
&VGG16 &65.9& 85.9&\textbf{44.8}\\ \cline{2-5}
 \hline
\multirow{2}{*}{STL-10}  & ResNet32
&100.0&35.0& \textbf{31.8}\\ 
 \cline{2-5}
&VGG16&64.1& 93.0& \textbf{27.0}\\ 
 \cline{2-5}
 \hline
\multirow{2}{*}{Tiny-ImageNet}  & ResNet32
 &100.0&96.3&\textbf{ 28.2}\\ 
 \cline{2-5}
&VGG16&69.7& 63.4& \textbf{54.4}\\ 
 \cline{2-5}

    \Xhline{1pt}
    \end{tabular}}\vspace{-2ex}
\end{table}

\noindent\textbf{Adaptive TA-LBF.}
Table~\ref{tab:adap-sam-asr} reports the defense results against adaptive TA-LBF, indicating the effectiveness of \sysname. 
Similarly, SDN becomes even more vulnerable than BASE under the adaptive TA-LBF attack. 
% For instance, on STL-10 and VGG16, \sysname reduces ASR to $26.8\%$, which is significantly lower than the ASR of BASE ($81.4\%$) and SDN ($89.9\%$). 
We analyze that compared with the basic SDN, \sysname still effectively mitigates the adaptive TA-LBF attack due to the \desdn. 

\begin{table}[h]\vspace{-2ex}
\centering
\caption{Evaluation results of ASR against adaptive TA-LBF. }
\label{tab:adap-sam-asr}
%%\vspace{-3ex}
\resizebox{0.75\linewidth}{!}{
    \begin{tabular}{c|c|c|c|c} \Xhline{1pt}
    \multirow{2}*{\textbf{Dataset}} &
    \multirow{2}*{\textbf{Model}}  & \multicolumn{3}{c}{\textbf{ASR (\%)}} 
    % & \multicolumn{3}{c}{\textbf{VGG16}} 
    \\\cline{3-5}
    &
     & \textbf{BASE}& \textbf{SDN}& \textbf{\sysname}   \\ \Xhline{1pt}
    \multirow{2}{*}{CIFAR-10}  & ResNet32
 &100.0&99.1& \textbf{60.8}\\  \cline{2-5}
&VGG16 &70.2&89.3& \textbf{50.3}\\ \cline{2-5}
 \hline
 \multirow{2}{*}{CIFAR-100}  & ResNet32
 &100.0&100.0& \textbf{26.4}\\  \cline{2-5}
&VGG16 &56.4& 78.2&\textbf{44.8}\\ \cline{2-5}
 \hline
\multirow{2}{*}{STL-10}  & ResNet32
&100.0&100.0& \textbf{10.2}\\ 
 \cline{2-5}
&VGG16&81.4&89.9& \textbf{26.8}\\ 
 \cline{2-5}
 \hline
\multirow{2}{*}{Tiny-ImageNet}  & ResNet32
 &100.0&100.0&\textbf{16.2}\\ 
 \cline{2-5}
&VGG16&51.8& 90.4& \textbf{15.0}\\ 
 \cline{2-5}
    \Xhline{1pt}
    \end{tabular}}\vspace{-2ex}
\end{table}

\begin{figure*}[!h]
    \centering
    \includegraphics[width=.95\textwidth,
    trim=0.cm 9.8cm 0.cm 0.1cm,clip]{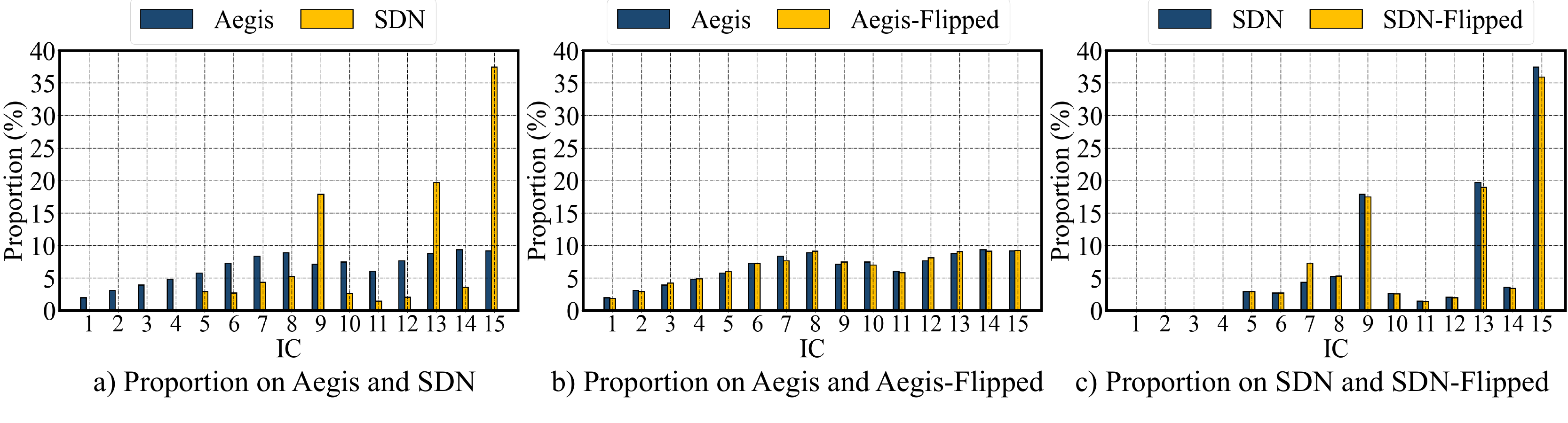} 
     \vspace{-4ex}
    \caption{The proportion of samples exit from different ICs or final layer (15 denotes the final layer) on CIFAR-100 and VGG16. Samples exit more uniformly in \sysname than SDN, even under BFA attacks (\sysname-Flipped and SDN-Flipped). }	
    \label{fig:distribution}\vspace{-3ex}
\end{figure*}

\noindent\textbf{Summary of evaluating adaptive TBT and TA-LBF.}
We analyze why \sysname is better than SDN against the adaptive TBT and TA-LBF is the \desdn scheme. 
Since SDN uses a static multi-exit mechanism, input samples have a stable exit pattern, i.e. most samples exit from a few fixed ICs.  
We give an example on CIFAR-100 and VGG16 to list the proportion of samples exiting pattern in \figurename~\ref{fig:distribution}. 
In \figurename~\ref{fig:distribution} (a), we observe that SDN lets more than $75\%$ samples exit from three exits, i.e. $IC_{9}$, $IC_{13}$, and $IC_{15}$.
Once the adversary modifies TBT and TA-LBF adaptively by including the loss for all ICs, its optimization process will utilize this exit pattern and locate the adaptive critical bits in these ICs to perform BFAs. 
\sysname adopts the \desdn mechanism to make input samples exit from all ICs uniformly. 
Even if the adversary adaptively attacks all ICs, the uniformly distributed exits will increase the attack cost (more bits to flip), thus enhancing security. 
Therefore, under the same $N_{b}$, \sysname significantly outperforms SDN.

We further consider whether the exit distributions of these defenses can be affected by the adaptive attack. Taking TA-LBF as an example, we denote the compromised SDN and \sysname as SDN-Flipped and \sysname-Flipped, respectively.
Figures~\ref{fig:distribution} (b) and (c) show that the original and flipped models of each method have similar exit distributions. 
Such results reveal that the distribution of \sysname is still uniform under the attack, giving the adversary no chances to utilize the exit pattern to locate the critical parameters of critical ICs. In contrast, the flipped SDN model exhibits the same exit distribution (i.e., vulnerability) as the original one.  
%than that of SDN even after attacks which gives the adversary no chance to utilize this exit pattern to locate the critical parameters of critical ICs. 

\noindent\textbf{Adaptive ProFlip.}
Table~\ref{tab:adap-pro-asr} shows the defense results for the adaptive ProFlip attack. 
% \sysname can still outperform SDN against the adaptive ProFlip. 
Compared with SDN and BASE, the ASR of \sysname is always significantly lower on all datasets and models.
Taking CIFAR-100 and ResNet32 as an example, the ASR of \sysname is $25.8\%$, while the ASR of SDN and BASE is $69.1\%$ and $89.8\%$ respectively.
We analyze the main reason:
with the optimization function for determining the critical parameters, the flipped bits are more likely to be concentrated in one layer.
Thus, the adversary cannot effectively affect ICs before the modified layer, making \sysname resilient against this attack. 
The reason why SDN has relatively good defense results in some cases also comes from this. 
However, SDN is much inferior to \sysname as it only uses a static early exit mechanism.
Such a static mechanism makes the number of affected ICs in the SDN greater than that in \sysname. Besides \desdn, our \rob contributes to the defense, which will be evaluated in Section~\ref{sec-abl-robust}.

\begin{table}[h]\vspace{-3ex}
\centering
\caption{Evaluation results of ASR against adaptive ProFlip. }
\label{tab:adap-pro-asr}
\resizebox{0.75\linewidth}{!}{
    \begin{tabular}{c|c|c|c|c} \Xhline{1pt}
    \multirow{2}*{\textbf{Dataset}} &
    \multirow{2}*{\textbf{Model}}  & \multicolumn{3}{c}{\textbf{ASR (\%)}} 
    % & \multicolumn{3}{c}{\textbf{VGG16}} 
    \\\cline{3-5}
    &
     & \textbf{BASE}& \textbf{SDN} & \textbf{\sysname}  \\ \Xhline{1pt}
    \multirow{2}{*}{CIFAR-10}  & ResNet32
 &96.9&74.2& \textbf{38.4}\\  \cline{2-5}
&VGG16 &88.2&79.1& \textbf{43.6}\\ \cline{2-5}
 \hline
 \multirow{2}{*}{CIFAR-100}  & ResNet32
 &89.8&69.1& \textbf{25.8}\\  \cline{2-5}
&VGG16 &80.0& 92.4&\textbf{33.7}\\ \cline{2-5}
 \hline
\multirow{2}{*}{STL-10}  & ResNet32
&77.4&57.8& \textbf{41.3}\\ 
 \cline{2-5}
&VGG16&87.2& 87.5& \textbf{34.5}\\ 
 \cline{2-5}
 \hline
\multirow{2}{*}{Tiny-ImageNet}  & ResNet32
 &99.1&64.4&\textbf{ 36.1}\\ 
 \cline{2-5}
&VGG16&88.2& 73.1& \textbf{40.8}\\ 
 \cline{2-5}

    \Xhline{1pt}
    \end{tabular}}
    \vspace{-2ex}
\end{table}

\noindent\textbf{Targeting shallow hidden layers.}
We further consider another adaptive attack, which focuses on shallow hidden layers to flip bits (denoted as \textbf{Shallow}).
Note that only ProFlip could be extended to achieve such an adaptive attack since TBT and TA-LBF must modify parameters connected to the target class, which are located in the last dense layer.

In particular, we modify ProFlip to choose critical bits among the first three hidden layers. Table~\ref{tab:adap-shallow} shows the results of Shallow. We observe that attacking shallow layers is not effective in bypassing \sysname.
Indeed, flipping bits in shallow layers cannot guarantee successful targeted attacks on the following hidden layers' outputs while \sysname lets samples randomly exit from arbitrary layers to limit ASR. 
Besides, targeting shallow layers may significantly decrease clean ACC ($6.2-27.8\%$), while the original ProFlip only degrades $2\%$. This makes such adaptive attacks easy to be detected.
%Particularly, Shallow degrades ACC by $6.2-27.8\%$, while original ProFlip only degrades $2\%$.} 

\begin{table}[!h]
\vspace{-2ex}
\centering
\caption{Evaluation results of attacking shallow layers. }
\label{tab:adap-shallow}
\resizebox{0.9\linewidth}{!}{
    \begin{tabular}{c|c|c|c|c|c} \Xhline{1pt}
    \multirow{2}*{\textbf{Dataset}} &
    \multirow{2}*{\textbf{Model}}  &\multicolumn{3}{c|}{\textbf{ASR (\%)}} &\multirow{2}*{\textbf{$\Delta$~ACC~(\%)}}
    % & \multicolumn{3}{c}{\textbf{VGG16}} 
    \\\cline{3-5}
    &
    & \textbf{BASE}
     & \textbf{\sysname}& \textbf{Shallow}
     \\ \Xhline{1pt}
    \multirow{2}{*}{CIFAR-10}  & ResNet32&96.9
 &38.4&40.5& \textbf{-18.7}\\  \cline{2-6}
&VGG16&88.2 &43.6&49.5& \textbf{-21.5}\\ \cline{2-5}
 \hline
 \multirow{2}{*}{CIFAR-100}  & ResNet32&89.8
 &25.8&22.8& \textbf{-6.2}\\  \cline{2-6}
&VGG16&80.0 &33.7& 37.0&\textbf{-8.8}\\ \cline{2-5}
 \hline
\multirow{2}{*}{STL-10}  & ResNet32&77.4
&41.3&48.4& \textbf{-23.1}\\ 
 \cline{2-6}
&VGG16&87.2&34.5& 51.7& \textbf{-8.9}\\ 
 \cline{2-6}
 \hline
\multirow{2}{*}{Tiny-ImageNet}  & ResNet32&99.1
 &36.1&30.3&\textbf{ -27.8}\\ 
 \cline{2-6}
&VGG16&88.2&40.8& 51.3& \textbf{-10.4}\\ 
 \cline{2-6}

    \Xhline{1pt}
    \end{tabular}}\vspace{-1ex}
\end{table}

\begin{figure*}[!h]
    \centering
    \includegraphics[width=.95\textwidth,
    trim=0.cm 9.8cm 0.cm 0.1cm,clip]{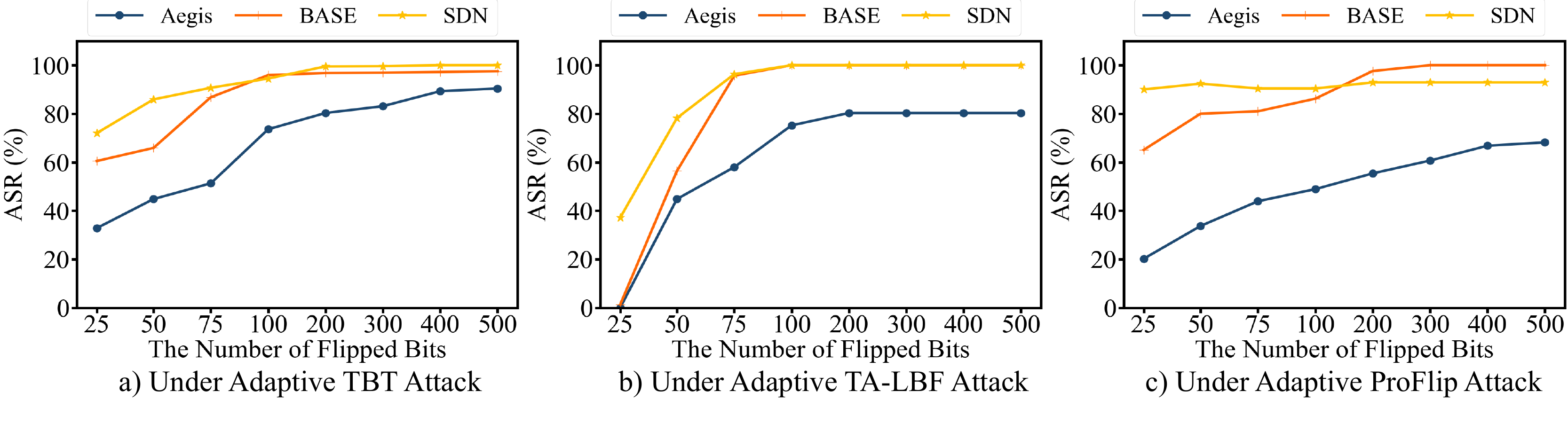} 
    \vspace{-4ex}
    \caption{On CIFAR-100 and VGG16, we compare \sysname with other defense methods under different values of $N_b$. Even the adversary significantly increases $N_{b}$, \sysname still outperforms others in the adaptive scenarios.}	
    \label{fig:asr_n_adap}\vspace{-3ex}
\end{figure*}

\noindent\textbf{Evaluation with more bits flipped.} 
We further evaluate the defense methods with up to 500 bits flipped for comprehensiveness. 
We conduct experiments on CIFAR-100 and VGG16 and adopt the adaptive TBT, TA-LBF, and ProFlip attacks. 
Results are shown in Figure~\ref{fig:asr_n_adap}.
Compared with existing defense methods, under the same value of $N_b$, we observe the ASR of \sysname is always the lowest against all attacks.

\subsection{Evaluation of Model Size}
~\label{section-modelsize}

%The extra model size increase introduced by \sysname is given in Table~\ref{tab:modelsize}. The model size increase for VGG16 is tiny but for ResNet is relatively larger which depends on datasets and structures. However, such a size increase is not a bottleneck for model deployment even on embedded devices like Nvidia Jetson Nano, (memory capacity is at GB level which is far more sufficient to support \sysname). Besides, \sysname allows samples to early-exit to significantly improve the inference efficiency with tiny ACC drops. 

\sysname can increase the model size during deployment. We evaluate the model size increase for all datasets and model structures used by us and find models ranging from 6.1MB to 97.6MB.
% as shown in Table~\ref{tab:modelsize}. 
We find that the size increase depends on the datasets and model structures.
Particularly, \sysname introduces a tiny increase for VGG16.
For example, on CIFAR-10, the size of VGG16 is 58.3MB and 65.6MB for BASE and \sysname, respectively.
In contrast,
ResNet is a small model so the increase is relatively larger.
For instance, also on CIFAR-10, 
the size of VGG16 is 1.9MB and 6.1MB for BASE and \sysname.
We emphasize that such a size increase is not a bottleneck for model deployment in practice. We find that for common embedded devices (e.g, Nvidia Jetson Nano), the memory capacity is usually at the GB level (far more sufficient to support \sysname). Besides, \sysname significantly improves the inference efficiency with almost no accuracy drop since most samples could early-exit from the network. 
Such benefits at the cost of acceptable size increase are very attractive for inference applications at the edge.

To validate the above points, we deploy our \sysname on two real-world widely-used edge devices: (1) Nvidia Jetson Nano with 4GB memory and 16GB storage; (2) Raspberry Pi 4 with 4GB memory and 32GB storage. The size increase is totally affordable for these two devices. We further evaluate the inference acceleration brought by \sysname.
We observe the average inference time is $46.1-59.4\%$ of the original model, which is a big improvement.

\subsection{Ablation Study on \rob}
\label{sec-abl-robust}
We verify the effectiveness of \rob. 
We compare \sysname with the setting without \rob, i.e., just \desdn.
Note that \rob aims to help ICs adapt to the adversarial scenario where bits in the layers attached by ICs are flipped. 
Therefore, we choose ProFlip to evaluate the effectiveness of \rob, as it can flip bits in the layers attached by ICs.
Table~\ref{tab:ablation-pro} shows that \rob effectively improves defense results. 
For the basic attacks, the ASR of \sysname is $3\%-13\%$ less than that of \desdn. 
%of \sysname is $19.2\%$ and $20.3\%$ on ResNet32 and VGG16 respectively, while the ASR of \desdn is much higher, i.e., $29.8\%$ and $28.7\%$, respectively. 
For the adaptive attacks, \sysname also performs better than \desdn by reducing $4\%-15\%$ of ASR.
Overall, we prove that \rob effectively contributes to mitigating targeted BFAs.

\begin{table}[!h]\vspace{-3ex}
\centering
\caption{Impact of \rob on basic and adaptive ProFlip.}
\label{tab:ablation-pro}
%%\vspace{-3ex}
\resizebox{0.9\linewidth}{!}{
    \begin{tabular}{c|c|c|c|c|c} \Xhline{1pt}
    \multirow{3}*{\textbf{Dataset}} & \multirow{3}*{\textbf{Model}} & \multicolumn{4}{c}{\textbf{ASR (\%)}} \\\cline{3-6}
    &  & \multicolumn{2}{c|}{\textbf{Basic ProFlip}} & \multicolumn{2}{c}{\textbf{Adaptive ProFlip}} \\\cline{3-6}
    & &\textbf{\desdn}& \textbf{\sysname}& \textbf{\desdn}& \textbf{\sysname}\\ \Xhline{1pt}
    \multirow{2}{*}{CIFAR-10}  & ResNet32 &24.1& \textbf{19.8} &45.1& \textbf{38.4}\\  \cline{2-6}
&VGG16& 33.7&\textbf{28.9}&49.4&\textbf{43.6} \\ \cline{2-6}
 \hline
 \multirow{2}{*}{CIFAR-100}  & ResNet32 &29.8& \textbf{19.2} &39.2& \textbf{25.8}\\  \cline{2-6}
&VGG16& 28.7&\textbf{20.3}&51.4&\textbf{33.7} \\ \cline{2-6}
 \hline
 \multirow{2}{*}{STL-10}  & ResNet32 &36.2& \textbf{33.9} &45.0& \textbf{41.3}\\  \cline{2-6}
&VGG16& 22.9&\textbf{18.7}&39.6&\textbf{34.5} \\ \cline{2-6}
 \hline
\multirow{2}{*}{Tiny-ImageNet}  & ResNet32
 &33.4&\textbf{20.1} &45.4&\textbf{36.1}\\  \cline{2-6}
&VGG16& 22.9& \textbf{15.6}&50.2& \textbf{40.8}\\  \cline{2-6}

    \Xhline{1pt}
    \end{tabular}}\vspace{-1ex}
\end{table}

\iffalse
\begin{table}[!h]
\centering
\caption{Impact of \rob on adaptive ProFlip.}
\label{tab:ablation-pro-adap}
%%\vspace{-3ex}
\resizebox{0.8\linewidth}{!}{
    \begin{tabular}{c|c|c|c} \Xhline{1pt}
    \multirow{2}*{\textbf{Dataset}} &
    \multirow{2}*{\textbf{Model}}  & \multicolumn{2}{c}{\textbf{ASR (\%)}} 
    % & \multicolumn{3}{c}{\textbf{VGG16}} 
    \\\cline{3-4}
    &
    & \textbf{\desdn} & \textbf{\sysname}\\ \Xhline{1pt}
 \multirow{2}{*}{CIFAR-100}  & ResNet32
 &39.2& \textbf{25.8}\\  \cline{2-4}
&VGG16 & 51.4&\textbf{33.7}\\ \cline{2-4}
 \hline
\multirow{2}{*}{Tiny-ImageNet}  & ResNet32
 &45.4&\textbf{36.1}\\ 
 \cline{2-4}
&VGG16& 50.2& \textbf{40.8}\\ 
 \cline{2-4}

    \Xhline{1pt}
    \end{tabular}}
\end{table}\fi

\vspace{-1ex}
\subsection{Attack Feasibility Analysis}\label{sec:feasibility}

%We use the TBT attack in a real system to validate the feasibility of BFAs. 
We validate the feasibility of BFAs (more specifically, TBT attack), 
% on a PC with GeForce GTX 1080 Ti GPU and Intel Haswell series processor (i5-4570) with AVX2 instruction set support. 
% The PC is configured with a dual-channel memory subsystem with one 4GB DDR3 DIMM (Hynix) in each channel. 
and use ResNet32 and Tiny-ImageNet as an example.
The main idea is to adopt the DeepSteal~\cite{DBLP:conf/sp/RakinCYF22} technique, which provides a memory massaging mechanism to realize Rowhammer~\cite{DBLP:conf/isca/KimDKFLLWLM14} attack, and is able to flip multiple bits within a 4KB page. 
This satisfies TBT's requirement, which targets one row of weights in the model's final layer and needs flipping multiple bits within a 4KB page.
This mechanism massages a page multiple times using memory swapping (the feature of swapping physical pages from DRAM to the disk swap space under memory pressure and then swapping them back when needed by the processor). Below we describe the details. 

\textit{Step 1: evicting victim pages.}
This step aims to evict the victim's pages from the main memory to swap space such that they can be relocated by the OS when they are accessed by the victim next time. To accomplish this, adversaries first allocate a large chunk of memory using \texttt{mmap} with the \texttt{MAP$\_$POPULATE} flag. 
This triggers the OS to evict other data (including the victim's pages) from the main memory to the swap space. 
Thus, we can occupy most of the physical memory space with victim pages stored in swap space. 

\textit{Step 2: releasing pages.}
This step aims to systematically release the occupied pages to enforce the desired relocation of the victim's pages. 
In detail, adversaries create a list of potential pages for the victim to occupy as aggressors during the attack.
At each round, adversaries choose a predetermined number of pages from the list and release the selected pages by calling \texttt{munmap}.

\textit{Step 3: deterministic relocation.}
This step aims to place victim pages in predetermined locations to create an appropriate memory layout for Rowhammer. 
It also ensures that the victim page location is known to adversaries so that they correlate flipped bits with exact data in the victim domain.
Adversaries follow DeepSteal~\cite{DBLP:conf/sp/RakinCYF22} to exploit the per-core page-frame cache structure to manipulate the operating system page allocation, which allows them to control where the victim pages are relocated. 

After the above steps, adversaries mount Rowhammer~\cite{DBLP:conf/isca/KimDKFLLWLM14} to flip bits in the victim pages placed in the appropriate locations. 
Since TBT targets one row of weights in a model's final layer which requires flipping multiple bits within a 4KB page, adversaries iterate the aforementioned steps until all target bits are flipped. 
Note that after each iteration, the weight page (with bit flips) will be swapped to the disk under memory pressure.
When this page is needed again, it is swapped back. 
As a consequence, bit flips will occur with this operation. When the weight page is swapped back, it has a high probability to be put into a new location in the memory where a different bit can be flipped. Then adversaries perform Rowhammer again to this page.  Adversaries iterate the entire process until all the required bits are flipped. 

%\noindent\textbf{Step 4: flipping bits.} This step aims to flip bits in the victim pages that are placed in the appropriate locations after Step 3. The adversary mounts Rowhammer to accomplish this.

% the memory massaging mechanism proposed in DeepSteal to improve Rowhammer.
% The mechanism massages a page multiple times by using memory swapping (the feature of swapping physical pages from DRAM to disk swap space under memory pressure and swapping them back when needed by the processor).
% In detail, we attack the target page of the weight that is allocated in the heap. 
% Under memory pressure, the weight pages (with bit flips) will be swapped to the disk. 
% When pages are needed again, they are swapped back. 
% As a result, the bit flips will come along in this operation. 
% Most likely when they are swapped back, they will be put into a new location in memory where a different bit can be flipped, and we do Rowhammer again to that page. 
% The whole process iterates until all target bits are flipped.

%as an example, and aim to flip 10 bits.
%The indices of the flipped bits by TBT are in Appendix~\ref{appendix-flippedbits}. 
%Results show the ASR on BASE is $77.8\%$. In contrast, the ASR on \sysname is much lower than that of BASE, i.e., only $2.0\%$. 
%The high ASR on the BASE model has validated the threat and feasibility of targeted BFAs.

\noindent\textbf{Evaluation results.}
Using the above technique, adversaries are able to flip 10 bits in the target model to achieve TBT. The index of the flipped bits can be found in Appendix~\ref{appendix-flippedbits}. For base models, flipping these bits can achieve an ASR of $77.8\%$. Now with our \sysname, the TBT attack can only get an ASR of 2.0\%. This confirms the effectiveness of \sysname. 

We further measure the attack cost. 
Flipping one bit takes about tens or even hundreds of seconds.
% In our experiment, flipping one bit takes about 300 seconds. 
%This is consistent with~\cite{yao2020deephammer}. 
Therefore, flipping more bits requires a much higher attack cost.  DeepHammer~\cite{yao2020deephammer} assumes that the maximum number of bits the adversary is allowed to flip is 24. To highlight the effectiveness of our defense, we consider a more powerful adversary who is allowed to flip $N_b=50$ bits (taking several hours) in Section~\ref{sec5-3}. Our \sysname is still effective against such an attack. Furthermore, we assume an unrealistically strong attack ($N_b=500$), and \sysname is still able to defeat it. Figures~\ref{fig:asr_n_nonadap} and ~\ref{fig:asr_n_adap} show the evaluation results for non-adaptive and adaptive attacks under this attack cost.

%Indeed, the time of flipping one bit varies and depends on the hardware. In our settings, flipping a bit takes nearly 600 seconds on average. We observe that the step of flipping bits takes $93\%$ of the time, while other steps take $7\%$ of the time. Therefore, the maximum number of flipped bits in DeepHammer~\cite{yao2020deephammer} is 24 which is reasonable. Recall our metric (i.e., $N_b=50$ in Section~\ref{sec5-3}), the time for flipping 50 bits is more than 8.3 hours which means this is extremely challenging for attackers. 
%In this case, \sysname still restricts ASR to a significantly low level. Furthermore, we relax $N_b$ to 500 in Figure~\ref{fig:asr_n_nonadap} (non-adaptive) and Figure~\ref{fig:asr_n_adap} (adaptive) to that show \sysname can still provide protection against targeted BFAs. 
%Besides, the maximum number of flipped bits in previous work~\cite{yao2020deephammer} is only 24, which is significantly less than 50.

% ~\wjl{
% }

% massage of a page multiple times at the cost of doing memory swapping. During attacks, we attack the anonymous page of the weight that is allocated in the heap. Under memory pressure, anonymous weight pages (with bit flips) will be swapped to the disk. When pages are needed again, they are swapped back. As a result, the bit flips will come along in this operation. Most likely when they are swapped back, they will be put into a new location in memory where a different bit can be flipped. There’s how you can do rowhammer again to that page. 

\section{Discussion and Future Work}\label{sec:discussion}

\noindent\textbf{Difference between \rob and adversarial training.}
Both adversarial training~\cite{madry2018towards} and our \rob aim to improve the model robustness by modifying model parameters but they are significantly different. 
Adversarial training aims to specifically defeat adversarial attacks by using perturbed samples generated from a clean model. 
Various previous work~\cite{DBLP:conf/iccv/RakinHF19,DBLP:conf/cvpr/HeRLCF20} have proved that directly using adversarial training cannot mitigate existing BFAs. 
In contrast, \rob improves robustness by considering a compromised model. 
By considering the effects brought by flipping critical bits of a target model, \rob can particularly increase the resistance against BFAs. 

\noindent\textbf{Floating-point DNN models.}
To the best of our knowledge, all state-of-the-art BFAs~\cite{DBLP:conf/iccv/RakinHF19,DBLP:conf/cvpr/RakinHF20,DBLP:conf/iccv/ChenFZK21,DBLP:conf/iclr/BaiWZL0X21} only focus on attacking quantized models. 
Therefore, we follow these existing works to evaluate our defense methods on quantized models in this paper. 
In fact, since \sysname is experimented in a non-intrusive fashion, protecting the floating-point DNN model is also feasible. 
We leave this experimentation as our future work.

% \noindent\textbf{The number of flipped bits.}
% We refer to the state-of-the-art Rowhammer attack~\cite{jattke2022blacksmith} and set $N_b$ to 50, as mentioned in Section~\ref{compare_metrics}.
% However, it is hard to justify this is the most appropriate choice, as Rowhammer is just one technique to achieve bit flips, and the attack cost is highly dependent on the hardware platforms.
% %bit-flip attacks are not equivalent to. In fact, bit-flip attacks have higher requirements than Rowhammer attacks, as they require precisely flipping any bits they want.
% Nevertheless, this setting allows us to guarantee all defense methods are fairly compared under the same $N_b$. Besides, to mitigate the potential effects brought by $N_b$, we also choose different values of $N_b$ to conduct more comprehensive evaluations, as shown in Figure~\ref{fig:asr_n_nonadap} and~\ref{fig:asr_n_adap}. Our method could still effectively mitigate BFAs as well as beat all other baselines under the same attack budget.

\noindent\textbf{Potential defense effects against untargeted BFAs.}
As indicated in Section~\ref{sec:theatModel}, mitigating untargeted BFAs is beyond the scope of this paper. 
However, we still evaluate \sysname against a state-of-the-art untargeted BFA~\cite{DBLP:conf/iccv/RakinHF19} for comprehensiveness. 
As shown in Appendix~\ref{appendix:untar-sec}, results show that \sysname could also mitigate untargeted BFA effectively by significantly increasing the attack cost (i.e. the number of flipped bits). 
%As a non-intrusive approach, \sysname is orthogonal to other defense methods (e.g. BIN~\cite{DBLP:conf/cvpr/HeRLCF20} or HashTAG~\cite{DBLP:conf/iccad/JavaheripiK21}) so it is feasible to improve these existing defenses by combining them with \sysname. 
We consider this as future work.

\section{Conclusion}
We propose \sysname, a novel mitigation methodology against targeted bit-flip attacks.
With a novel design of \desdn, we randomly select ICs for inference, enabling input samples to early-exit from them and effectively obfuscate the adversary.
We further propose \rob to improve IC robustness. 
We conduct extensive experiments with four mainstream datasets and two DNN structures
to show that \sysname can mitigate various state-of-the-art targeted attacks as well as their adaptive versions, and significantly outperform existing defense methods.

\section*{Acknowledgments}
This work is supported by the National Key R\&D Program
of China (2022YFB3105202), National Natural Science Foundation of China (62106127, 62132011, 61972224), NSFOCUS (2022671026), Singapore Ministry of Education (MOE) AcRF Tier 2 MOE-T2EP20121- 0006, and Ant Group through CCF-Ant Innovative Research Program No. RF2021002.

\bibliography{reference}
\bibliographystyle{plain}

\clearpage
% \section*{Appendix}
\appendix

\section{Analysis of Hyperparameters }
% \subsection*{Hyperparameters of Training Models}
\label{appendix:hyper_sensitivity}
Tuning $\tau$ and $q$ are not related to any specific attacks. 
We tune these two hyperparameters for two goals on clean models only, i.e. (1) making early-exits uniformly distributed to prevent the attacker from targeting only those popular exits; (2) maintaining high ACC on benign samples.
Our results are stable to hyperparameters that meet these two requirements. 
We evaluate the sensitivity of these two hyperparameters on CIFAR-10 and ResNet32. 

In particular, we first fix $q=3$ and set $\tau$ to $0.93$, $0.95$, $0.97$, respectively, to observe the sensitivity of $\tau$ on mitigating the attacks.
Then, we fix $\tau=0.95$, and set $q$ to $2$, $3$, $4$, respectively, to observe the sensitivity of $q$. 
Note that, these chosen values of $\tau$ and $q$ could make early-exits uniformly distributed, as well as restrict the degradation of accuracy below $1.35\%$.
Table~\ref{tab:tunehyper-tau} and~\ref{tab:tunehyper-q} present the mitigation results considering different $\tau$ and $q$.
For all three attacks, under both non-adaptive and adaptive settings, the ASR variance is below $2\%$.
Such results reveal that the effectiveness of \sysname is stable if the chosen values of $\tau$ and $q$ meet the two goals.
In contrast, if we violate the two goals to tune the hyperparameters, the mitigation results should be hugely affected.
For example, if we randomly set $\tau=0.7$ and $q=8$, most samples would early exit from the shallow layers. 
In this case, the ACC significantly degrades and the adaptive adversary could focus on the popular exits to improve the ASR.

\begin{table}[!htbp]\vspace{-3ex}
\centering
\caption{Evaluation results of the sensitivity on $\tau$. }
\label{tab:tunehyper-tau}
%%\vspace{-3ex}
\resizebox{0.75\linewidth}{!}{
    \begin{tabular}{c|c|c|c} \Xhline{1pt}
    \multirow{2}*{\textbf{Attacks}} &
    \multirow{2}*{\textbf{$\tau$}}  & \multicolumn{2}{c}{\textbf{ASR (\%)}} 
    % & \multicolumn{3}{c}{\textbf{VGG16}} 
    \\\cline{3-4}
    & & \textbf{Non-adaptive}& \textbf{Adaptive}   \\ \Xhline{1pt}
    \multirow{3}{*}{TBT}  & 0.93
 &19.1&30.2\\  \cline{2-4}
&0.95 &19.9&31.1\\ \cline{2-4}
&0.97 &21.1&31.6\\ \cline{1-4}
\multirow{3}{*}{TA-LBF}  & 0.93
 &5.1&59.6\\  \cline{2-4}
&0.95 &6.3&60.8\\ \cline{2-4}
&0.97 &7.0&60.2\\ \cline{1-4}
\multirow{3}{*}{ProFlip}  & 0.93
 &18.4&38.5\\  \cline{2-4}
&0.95 &19.8&38.4\\ \cline{2-4}
&0.97 &20.3&38.0\\ \cline{2-4}
 \hline

    \Xhline{1pt}
    \end{tabular}}
\end{table}

\begin{table}[!htbp]\vspace{-3ex}
\centering
\caption{Evaluation results of the sensitivity on $q$. }
\label{tab:tunehyper-q}
%%\vspace{-3ex}
\resizebox{0.75\linewidth}{!}{
    \begin{tabular}{c|c|c|c} \Xhline{1pt}
    \multirow{2}*{\textbf{Attacks}} &
    \multirow{2}*{\textbf{$q$}}  & \multicolumn{2}{c}{\textbf{ASR (\%)}} 
    % & \multicolumn{3}{c}{\textbf{VGG16}} 
    \\\cline{3-4}
    & & \textbf{Non-adaptive}& \textbf{Adaptive}   \\ \Xhline{1pt}
    \multirow{3}{*}{TBT}  & 2
 &19.5&29.2\\  \cline{2-4}
&3 &19.9&31.1\\ \cline{2-4}
&4 &20.0&31.3\\ \cline{1-4}
\multirow{3}{*}{TA-LBF}  & 2
 &6.5&59.8\\  \cline{2-4}
&3 &6.3&60.8\\ \cline{2-4}
&4 &4.7&61.2\\ \cline{1-4}
\multirow{3}{*}{ProFlip}  & 2
 &19.4&38.3\\  \cline{2-4}
&3 &19.8&38.4\\ \cline{2-4}
&4 &19.0&38.6\\ \cline{2-4}
 \hline

    \Xhline{1pt}
    \end{tabular}}\vspace{-2ex}
\end{table}

\section{Hyperparameters of Attacks}
% \subsection*{Hyperparameters of Attacks}
\label{appendix-attacks}
\noindent\textbf{TBT Hyperparameters.} We follow the same settings  in~\cite{DBLP:conf/cvpr/RakinHF20}. 
In particular,
there are three hyperparameters: $t$, $w_b$, and TAP.
$t$ is the attack target class. 
We randomly select $t$ in each independent experiment and employ SGD with a learning rate of 0.5 for 200 epochs.
% We also randomly sample 256 samples from the test sets to train the trigger.
$w_b$ is the number of weights to be flipped, and we set $w_b=10$.
TAP is the percentage of the area of the trigger trained by the attacker to the area of the input image, and we set the TAP to 9.76\%.

% \section{Parameter settings of attacks.}
% \subsection{TA-LBF attack.}

\noindent\textbf{TA-LBF Hyperparameters.}
We follow the same settings in ~\cite{DBLP:conf/iclr/BaiWZL0X21}.
There are three hyperparameters: $N$, $k$, and $\lambda$.
$N$ controls the effect on model accuracy and here we choose a fixed number 128 for $N$.
$k$ limits the number of flipped bits.
Here we set the initial $k$ as 5 and the maximum searching iteration for $k$ as 6 for each attack.
$\lambda$ is a trade-off hyperparameter that keeps a balance between the attack effectiveness and the model utility.
Here we set the initial $\lambda$ as 100 and the maximum searching times as 8 for each attack.
Note that for each dataset, we choose 1000 samples (each with a pre-assigned targeted label different from the ground-truth label) for TA-LBF attack. 
For each of the 10 classes in CIFAR-10 and STL-10, we randomly select 100 validation images from the other 9 classes. 
For CIFAR-100 and Tiny-ImageNet, we randomly choose 50 target classes, and for each of them randomly select 20 validation images from the other classes.

\noindent\textbf{ProFlip Hyperparameters.}
We follow the same settings in~\cite{DBLP:conf/iccv/ChenFZK21}.
In particular,
there are two hyperparameters: $t$, $k$, and TAP.
$t$ controls the attack target class. 
We randomly select $t$ in each independent experiment.
In order to reduce the computational complexity, in each iteration, ProFlip determines the critical bits by traversing $k$ points between the maximum and minimum values of the parameters. 
We choose $k=20$ for CIFAR-10 and CIFAR-100, and $k=10$ for STL-10 and Tiny-ImageNet.
TAP is the percentage of the area of the trigger trained by the attacker to the area of the input image, and we set the TAP to 9.76\%.

\section{Visualization of Triggers}
\label{appendix:vis_trigger}
\begin{figure}[!h]
    \centering
    
 \includegraphics[width=.99\textwidth,trim=0.0cm 10.5cm 5.2cm 0.0cm,clip
    ]{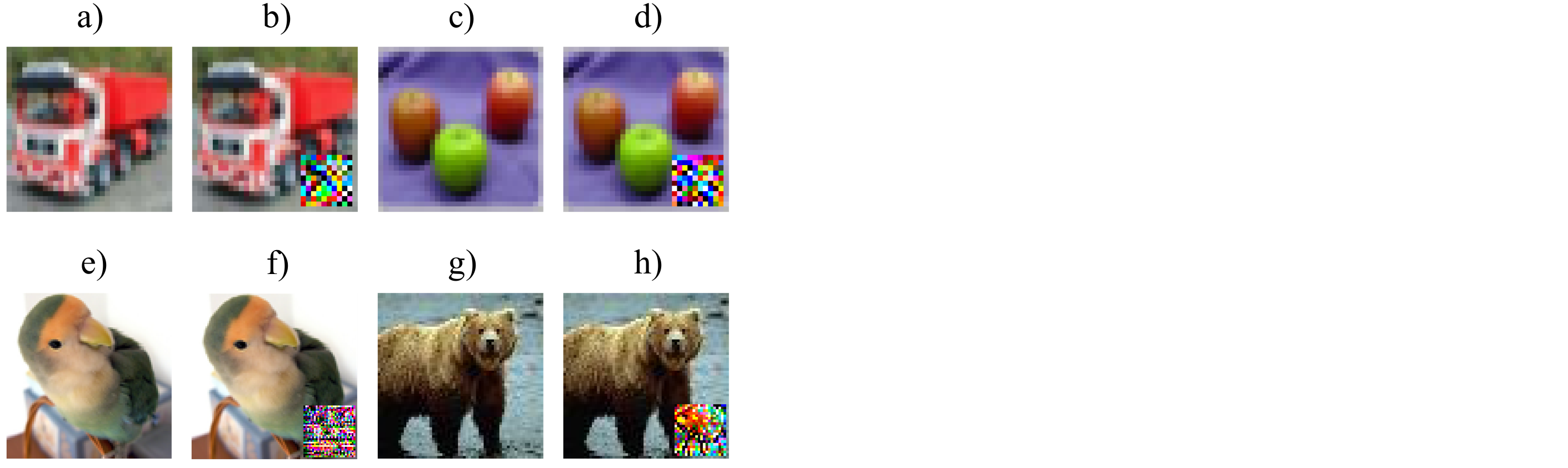} 
    %%%\vspace{-0.2cm}
    \caption{Examples of clean and triggered samples for TBT: a) and b) for CIFAR-10, c) and d) for CIFAR-100, e) and f) for STL-10, g) and h) for Tiny-ImageNet.}
    \label{fig:app_tbt}
    %%\vspace{-3ex}
% \setlength{\belowcaptionskip}{-1.5cm}
\end{figure}

\begin{figure}[!h]
    \centering
    
 \includegraphics[width=.99\textwidth,trim=0.0cm 10.5cm 5.2cm 0.0cm,clip
    ]{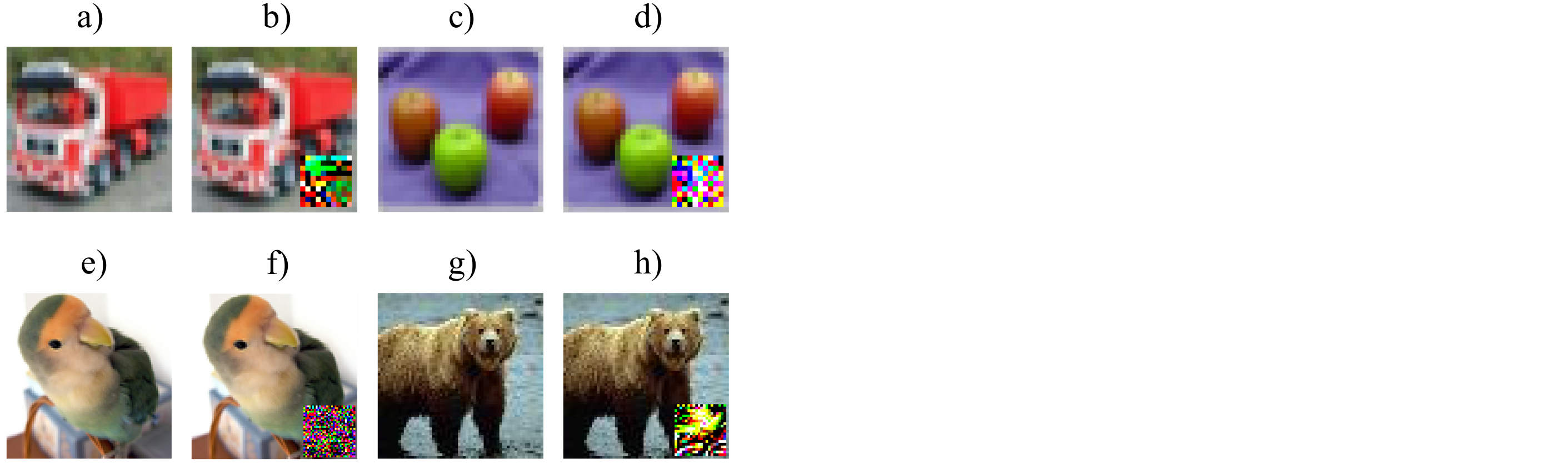} 
    %%%\vspace{-0.2cm}
    \caption{Examples of clean and triggered samples for ProFlip: a) and b) for CIFAR-10, c) and d) for CIFAR-100, e) and f) for STL-10, g) and h) for Tiny-ImageNet.}
    \label{fig:app_pro}
    \vspace{-2ex}
\end{figure}

We give several examples to show the triggers generated by TBT and ProFlip.
Both traditional backdoor attacks~\cite{DBLP:conf/ccs/ShenJ0LCSFYW21,DBLP:conf/ccs/YaoLZZ19} and these two targeted BFAs generate triggers but they are significantly different. 
Traditional backdoor attacks tend to specify a pre-determined trigger.
In contrast, TBT and ProFlip train a trigger from scratch to activate some specific neurons in DNNs.
Thus, we can observe the triggers in Figure~\ref{fig:app_tbt}, ~\ref{fig:app_pro} have no specific style.

% The main reason for the effectiveness of our method is that: the flipped bits are specific for attacking the final layer, but not ICs. 

\section{Mitigating Untargeted Attacks}
\label{appendix:untar-sec}
Although \sysname is not designed to mitigate untargeted BFAs, we still evaluate it against untargeted BFAs. 
We use the state-of-the-art untargeted BFA~\cite{DBLP:conf/iccv/RakinHF19}.
We reproduce the untargeted attack method by directly using their source code with recommended parameters.
The untargeted attack aims to degrade model accuracy to random guess.
For different datasets, we make these accuracy explicit as follows. 

\begin{packeditemize}
\item 
On CIFAR-10, each image of the dataset belongs to one of 10 classes, and the accuracy equivalent to random guess is $10\%$.
Therefore, on this dataset, attackers aim to degrade the accuracy of models to $10\%$.

\item 
On CIFAR-100, each image of the dataset belongs to one of 100 classes, and the accuracy equivalent to random guess is $1\%$.
Therefore, on this dataset, attackers aim to degrade the accuracy of models to $1\%$.

\item 
On STL-10, each image of the dataset belongs to one of 10 classes, and the accuracy equivalent to random guess is $10\%$.
Therefore, on this dataset, attackers aim to degrade the accuracy of models to $10\%$.

\item 
On Tiny-ImageNet, each image of the dataset belongs to one of 200 classes, and the accuracy equivalent to random guess is $0.5\%$.
Therefore, on this dataset, attackers aim to degrade the accuracy of models to $0.5\%$.

\end{packeditemize}

\begin{table}[!h]
\centering
\caption{Evaluation results of ASR against untargeted attack.}
\label{tab:untar}
%%\vspace{-3ex}
\resizebox{0.8\linewidth}{!}{
    \begin{tabular}{c|c|c|c} \Xhline{1pt}
    \multirow{2}*{\textbf{Dataset}} &
    \multirow{2}*{\textbf{Model}}  & \multicolumn{2}{c}{\textbf{$N_{b}$}} 
    % & \multicolumn{3}{c}{\textbf{VGG16}} 
    \\\cline{3-4}
    &
     & \textbf{BASE}& \textbf{\sysname}\\ \Xhline{1pt}
      \multirow{2}{*}{CIFAR-10}  & ResNet32
 &20& \textbf{96}\\  \cline{2-4}
&VGG16& 24&\textbf{74} \\ \cline{2-4}
 \hline
 \multirow{2}{*}{CIFAR-100}  & ResNet32
 &7& \textbf{269}\\  \cline{2-4}
&VGG16& 39&\textbf{79} \\ \cline{2-4}
 \hline
  \multirow{2}{*}{STL-10}  & ResNet32
 &13& \textbf{39}\\  \cline{2-4}
&VGG16& 117&\textbf{3068} \\ \cline{2-4}
 \hline
\multirow{2}{*}{Tiny-ImageNet}  & ResNet32
 &17&\textbf{64}\\ 
 \cline{2-4}
&VGG16& 68& \textbf{143}\\ 
 \cline{2-4}

    \Xhline{1pt}
    \end{tabular}}\vspace{-2ex}
\end{table}

Following previous work~\cite{DBLP:conf/cvpr/HeRLCF20}, we evaluate the number of flipped bits $N_{b}$ required to make the untargeted attack successful, i.e., degrading model ACC to the random guess level. 
The higher $N_{b}$, the better the defense.

Table~\ref{tab:untar} 
shows the results under non-adaptive and adaptive scenarios respectively. We can observe that \sysname could effectively mitigate the untargeted attack, as we significantly increase the $N_{b}$ for attacks. 
n different datasets and models, our \sysname can effectively improve $N_{b}$ by $2.0\sim 38.4\times$.
On CIFAR-100 dataset, taking ResNet32 as an example, attackers need to flip 269 bits to attack models protected by our \sysname,
but only need to flip 7 bits to attack BASE.
Thus, \sysname improves 
model resistance by  $\sim 38.4\times$.

\section{Flipped Bits}
~\label{appendix-flippedbits}
On Tiny-ImageNet and ResNet32, we flip 10 bits in the final dense layer, and the positions of bits to flip are listed here: 
[(2, 1, 0), (2, 15, 0), (2, 157, 7), (2, 166, 7), (2, 169, 7), (2, 189, 7), (2, 218, 7), (2, 238, 7), (2, 253, 6), (2, 255, 2)].

The shape of the final dense layer is [200, 256].
(2, 1, 0) denotes: flip the 0th bit of parameter [2, 1]. 
(2, 15, 0) denotes: flip the 0th bit of parameter [2, 15].
……
(2, 157, 7) denotes: flip the 7th bit of the parameter [2, 157].

Given a binary, e.g., 10001000 (decimal: 136), the index of bits is [7th, 6th, 5th, 4th, 3rd, 2nd, 1st, 0th]

\end{document}